\documentclass[twocolumn]{aastex62}
\usepackage{natbib}
\usepackage{color}
\usepackage{mathptmx}
\usepackage{amsmath}
\usepackage{url}
\usepackage{multirow}
\usepackage{verbatim}
\def\msun{\,{\rm M_\odot}}
\def\Rvir{\,R_{\rm vir}}
\def\Rvirtext{$\,R_{\rm vir}\,$}
\def\redshift #1{$z=#1$}
\def\code #1{{\sc #1}}
\def\zsun{\,{Z_\odot}}
\usepackage{numprint}
\usepackage{graphicx}
\usepackage{refcount}
\usepackage{etoolbox}
\usepackage{blindtext}
\usepackage{graphicx}
\usepackage{everypage}
\usepackage{environ}
\graphicspath{{./}{figures/}}
\received{August 1, 2023}
\revised{\today}
\accepted{\today}

\shorttitle{{\it AGORA} Comparison. VI: The Circumgalactic Medium}
\shortauthors{{\it AGORA} Collaboration et al.}

\newcounter{abspage}

\makeatletter
\newcommand{\newSFPage}[1]
  {\global\expandafter\let\csname SFPage@#1\endcsname\null}

\NewEnviron{SidewaysFigure}{
\begin{figure*}[p]
\protected@write\@auxout{\let\theabspage=\relax}
  {\string\newSFPage{\theabspage}}%
\ifdim\textwidth=\textheight
  \rotatebox{90}{\parbox[c][\textwidth][c]{\linewidth}{\BODY}}%
\else
  \rotatebox{90}{\parbox[c][\textwidth][c]{\textheight}{\BODY}}%
\fi
\end{figure*}
}

\AddEverypageHook{
  \ifdim\textwidth=\textheight
    \stepcounter{abspage}
  \else
    \@ifundefined{SFPage@\theabspage}{}{\global\pdfpageattr{/Rotate 0}}%
    \stepcounter{abspage}%
    \@ifundefined{SFPage@\theabspage}{}{\global\pdfpageattr{/Rotate 90}}%
  \fi}
\makeatother

\begin{document}
\title{The {\it AGORA} High-resolution Galaxy Simulations Comparison Project. VI: Similarities and Differences in the Circumgalactic Medium}

\author[0000-0001-9695-4017]{Clayton Strawn}
\affil{Department of Physics, University of California at Santa Cruz, Santa Cruz, CA 95064, USA; \rm{\href{mailto:cjstrawn@ucsc.edu}{cjstrawn@ucsc.edu}}}

\author[0000-0002-6299-152X]{Santi Roca-F\`{a}brega}
\altaffiliation{Code leaders}
\affil{Lund Observatory, Division of Astrophysics, Department of Physics, Lund University, SE-221 00 Lund, Sweden; \rm{\href{mailto:santi.roca_fabrega@fysik.lu.se}{santi.roca\_fabrega@fysik.lu.se}}}
\affil{Departamento de F\'{i}sica de la Tierra y Astrof\'{i}sica, Facultad de Ciencias F\'{i}sicas, Plaza Ciencias, 1, 28040 Madrid, Spain}

\author[0000-0001-5091-5098]{Joel R. Primack}
\affil{Department of Physics, University of California at Santa Cruz, Santa Cruz, CA 95064, USA; \rm{\href{mailto:joel@ucsc.edu}{joel@ucsc.edu}}}

\author[0000-0003-4464-1160]{Ji-hoon Kim}
\altaffiliation{Code leaders}
\affiliation{Seoul National University Astronomy Research Center, Seoul 08826, Korea; \rm{\href{mailto:mornkr@snu.ac.kr}{mornkr@snu.ac.kr}}}
\affiliation{Center for Theoretical Physics, Department of Physics and Astronomy, Seoul National University, Seoul 08826, Korea}

\author{Anna Genina}
\altaffiliation{Code leaders}
\affil{Max-Planck-Institut f\"{u}r Astrophysik, Karl-Schwarzschild-Str. 1, D-85748, Garching, Germany}

\author[0000-0002-4687-4948]{Loic Hausammann}
\altaffiliation{Code leaders}
\affil{Institute of Physics, Laboratoire d'Astrophysique, \'{E}cole Polytechnique F\'{e}d\'{e}rale de Lausanne (EPFL), CH-1015 Lausanne, Switzerland}
\affil{TS High Performance Computing, Eidgen\"ossische Technische Hochschule Z\"urich (ETHZ), 8092 Z\"urich, Switzerland}

\author[0000-0002-7820-2281]{Hyeonyong Kim}
\altaffiliation{Code leaders}
\affiliation{Center for Theoretical Physics, Department of Physics and Astronomy, Seoul National University, Seoul 08826, Korea}
\affiliation{Department of Aerospace Engineering, Seoul National University, Seoul 08826, Korea}

\author{Alessandro Lupi}
\altaffiliation{Code leaders}
\affil{DiSAT, Universit\`a degli Studi dell'Insubria, via Valleggio 11, I-22100 Como, Italy}
\affil{Dipartimento di Fisica ``G. Occhialini'', Universit\`a degli Studi di Milano-Bicocca, I-20126 Milano, Italy}

\author[0000-0001-7457-8487]{Kentaro Nagamine}
\altaffiliation{Code leaders}
\affiliation{Department of Earth and Space Science, Graduate School of Science, Osaka University, Toyonaka, Osaka, 560-0043, Japan}
\affiliation{Kavli IPMU (WPI), University of Tokyo, 5-1-5 Kashiwanoha, Kashiwa, Chiba, 277-8583, Japan}
\affiliation{Department of Physics \& Astronomy, University of Nevada Las Vegas, Las Vegas, NV 89154, USA}

\author[0000-0002-3764-2395]{Johnny W. Powell}
\altaffiliation{Code leaders}
\affil{Department of Physics, Reed College, Portland, OR 97202, USA}

\author{Yves Revaz}
\altaffiliation{Code leaders}
\affil{Institute of Physics, Laboratoire d'Astrophysique, \'{E}cole Polytechnique F\'{e}d\'{e}rale de Lausanne (EPFL), CH-1015 Lausanne, Switzerland}

\author{Ikkoh Shimizu}
\altaffiliation{Code leaders}
\affil{Shikoku Gakuin University, 3-2-1 Bunkyocho, Zentsuji, Kagawa, 765-8505, Japan}

\author{H\'{e}ctor Vel\'{a}zquez}
\altaffiliation{Code leaders}
\affil{Instituto de Astronom\'{i}a, Universidad Nacional Aut\'{o}noma de M\'{e}xico, A.P. 70-264, 04510, Mexico, D.F., Mexico}

\author[0000-0002-5969-1251]{Tom Abel}
\affil{Kavli Institute for Particle Astrophysics and Cosmology, Stanford University, Stanford, CA 94305, USA}
\affil{Department of Physics, Stanford University, Stanford, CA 94305, USA}
\affil{SLAC National Accelerator Laboratory, Menlo Park, CA 94025, USA}

\author{Daniel Ceverino}
\affil{Universidad Aut\'{o}noma de Madrid, Ciudad Universitaria de Cantoblanco, E-28049 Madrid, Spain}
\affil{CIAFF, Facultad de Ciencias, Universidad Aut\'{o}noma de Madrid, E-28049 Madrid, Spain}

\author{Bili Dong}
\affil{Department of Physics, Center for Astrophysics and Space Sciences, University of California at San Diego, La Jolla, CA 92093, USA}

\author[0000-0002-9144-1383]{Minyong Jung}
\affiliation{Center for Theoretical Physics, Department of Physics and Astronomy, Seoul National University, Seoul 08826, Korea}

\author[0000-0001-5510-2803]{Thomas R. Quinn}
\affil{Department of Astronomy, University of Washington, Seattle, WA 98195, USA}

\author[0000-0002-4639-5285]{Eun-jin Shin}
\affiliation{Center for Theoretical Physics, Department of Physics and Astronomy, Seoul National University, Seoul 08826, Korea}

\author[0000-0002-8638-1697]{Kirk S.~S.~Barrow}
\affiliation{Department of Astronomy, University of Illinois at Urbana-Champaign, Urbana, IL 61801, USA}

\author{Avishai Dekel}
\affil{Center for Astrophysics and Planetary Science, Racah Institute of Physics, The Hebrew University, Jerusalem 91904, Israel}

\author[0000-0003-4597-6739]{Boon Kiat Oh}
\affiliation{Department of Physics, University of Connecticut, U-3046, Storrs, CT 06269, USA}
\affiliation{Center for Theoretical Physics, Department of Physics and Astronomy, Seoul National University, Seoul 08826, Korea}

\author{Nir Mandelker}
\affil{Center for Astrophysics and Planetary Science, Racah Institute of Physics, The Hebrew University, Jerusalem 91904, Israel}

\author{Romain Teyssier}
\affil{Department of Astrophysical Sciences, Princeton University, Princeton, NJ 08544, USA}

\author[0000-0002-3817-8133]{Cameron Hummels}
\affiliation{TAPIR, California Institute of Technology, Pasadena, CA 91125, USA}

\author{Soumily Maji}
\altaffiliation{High School students who worked with the Collaboration through\\ the UC Santa Cruz Science Internship Program (SIP).}
\affil{Fremont High School, Sunnyvale, CA 94087, USA}

\author{Antonio Man}
\altaffiliation{High School students who worked with the Collaboration through\\ the UC Santa Cruz Science Internship Program (SIP).}
\affil{Green Valley High School, Henderson, NV 89014, USA}

\author{Paul Mayerhofer}
\altaffiliation{High School students who worked with the Collaboration through\\ the UC Santa Cruz Science Internship Program (SIP).}
\affil{Redwood High School, Larkspur, CA 94939, USA}

\author{the {\it AGORA} Collaboration}
\affiliation{\rm \url{http://www.AGORAsimulations.org}}
\affiliation{\rm The authors marked with * as code leaders contributed to the article by leading the effort within each code group to perform and analyze simulations.} 




\begin{abstract}
We analyze the circumgalactic medium (CGM) for eight commonly-used cosmological codes in the AGORA collaboration. The codes are calibrated to use identical initial conditions, cosmology, heating and cooling, and star formation thresholds, but each evolves with its own unique code architecture and stellar feedback implementation. Here we analyze the results of these simulations in terms of the structure, composition, and phase dynamics of the CGM. We show properties such as metal distribution, ionization levels, and kinematics are effective tracers of the effects of the different code feedback and implementation methods, and as such can be highly divergent between simulations. This is merely a fiducial set of models, against which we will in the future compare multiple feedback recipes for each code. Nevertheless, we find that the large parameter space these simulations establish can help disentangle the different variables that affect observable quantities in the CGM, e.g. showing that abundances for ions with higher ionization energy are more strongly determined by the simulation's metallicity, while abundances for ions with lower ionization energy are more strongly determined by the gas density and temperature.

\end{abstract}



\defcitealias{kim_agora_2013}{Paper I}
\defcitealias{kim_agora_2016}{Paper II}
\defcitealias{roca-fabrega_agora_2021}{Paper III}
\defcitealias{roca-fabrega_inprep}{Paper IV}
\defcitealias{jung_inprep}{Paper V}

\section{Introduction}\label{sec:intro}
The circumgalactic medium, or CGM, is usually defined as the baryonic matter which resides within the virial radius \Rvirtext but outside the galaxy ``boundary'', for which a number of different definitions exist. We will use the value $0.15 \Rvir$, corresponding to the expected size of the galaxy disk, though this is significantly larger than other common boundary definitions like the half-mass radius \cite[see][Appendix A for a discussion of the evolution of half-mass radii in simulations]{rohr_galaxy-halo_2022}. This gas is essential for any meaningful understanding of the long-term growth and evolution of galaxies, because any gas which flows into or out of a visible galaxy, for use in star formation within a galaxy disk or metal pollution of the intergalactic medium (IGM), has to pass through this region \citep{woods_role_2014}. In transit, it is caught up in a web of dynamical forces operating in a physical regime which is quite distinct from that of the other populations of gas in the universe, such as the interstellar medium (ISM), affected by active galactic nuclei (AGN), star formation, and dynamical perturbuations due to clumps, or gas within the extremely low-density intergalactic medium (IGM), dominated by cosmological effects. A summary of the current state of the theory of CGM dynamics can be found in \cite{faucher-giguere_key_2023}, and references therein.

Interest in the CGM has grown considerably in recent years, as the significance of this region has become more apparent to the galaxy formation community and more data has become available \citep[See][and references therein, for a summary of the observational picture]{tumlinson_circumgalactic_2017}. Due to its low density, the CGM is very difficult to see in emission line mapping, with the exceptions being H I emission \citep{zhang_hydrogen_2016,cai_evolution_2019}, which is unfortunately not a very good tracer of higher temperature gas, and metal line emission which is usually only possible in very nearby galaxies at \redshift{0} \citep{howk_project_2017,li_circum-galactic_2017}. Instead, the CGM tends to be observed in absorption against bright background sources, generally quasar spectra. In the last decade, there has been a tremendous increase in the amount of observational data available due to the development of improved space-based and ground-based telescopes, including the groundbreaking COS-Halos survey \citep[e.g.][]{tumlinson_large_2011,werk_cos-halos_2013,werk_cos-halos_2014} and an expanding number of new and larger samples, e.g. KBSS \citep{rudie_column_2019}, CASBaH \citep{prochaska_cos_2019,burchett_cos_2019}, CUBS \citep{chen_cosmic_2020}, and CGM$^2$ \citep{wilde_cgm2_2021,tchernyshyov_cgm2_2022}, among many others.

Because absorption line spectroscopy requires a coincidence between background sources and foreground galaxies, it is very rare to get multiple sightlines of data around any single galaxy, though this is possible, either through coincidence \citep[e.g.][]{keeney_hstcos_2013}, or through exploiting the effects of strong lensing by the foreground halo to see the same background object in multiple places \citep[e.g.][]{ellison_sizes_2004,okoshi_multiple_2019}. It is especially challenging because separate imaging and spectroscopy tools are needed to analyze the hosting galaxy system and the quasar sightline. Together this means that there is still significant uncertainty regarding the physical state of gas in the region, and that maximal information needs to be extracted from each line of sight.

As a rule, the CGM is highly ionized, and much of the interpretation of the physical state of gas, therefore, comes from interpreting absorption lines from ionized metals, in particular their column density, Doppler broadening, and kinematic alignment with one another. Metal lines have the advantage of relatively low line confusion with the Lyman alpha forest, and they are more likely than hydrogen to be in the linear regime and not saturated. Ionized metal densities can be a very good test of the physical state and evolution of the CGM because they are very sensitive to multiple variables, all of which can vary continuously. The number density of an element $X$ in ionization state $i$ is 
\begin{equation}
    n_{X_i} = A_X \cdot n \cdot Z \cdot f_{X_i},
    \label{eq:factors}
\end{equation}
where $A_X$ is the fractional abundance of element $X$ per metallicity unit, $n$ is the number density of gas, $Z$ is the overall metallicity in that parcel, and finally $f_{X_i}$ is the fraction of the element $X$ in state $i$, at the parcels given temperature and density. In this work, $A_X$ is assumed to be the constant solar abundance value, e.g. the number of carbon, oxygen, etc. nuclei for each hydrogen nucleus (at $Z=\zsun$), which are taken from \code{Cloudy} documentation \citep{ferland_2013_2013}.\footnote{In the real Universe, $A_X$ would affected by differences in elemental metal production from different sources, such as Type Ia SNe producing more iron-peak elements, and Type II SNe producing more alpha elements, but since not all AGORA codes track these species independently it was decided to use the solar ratios for all.}

This extreme sensitivity to multiple variables makes the CGM an interesting area of focus for the AGORA (Assembling Galaxies of Resolved Anatomy) code comparison project, whose earlier simulations are shown in \cite{kim_agora_2013,kim_agora_2016}, hereafter Papers I and II, respectively. This large international collaboration of leading simulation code researchers is dedicated to examining the convergence or divergence of different simulation codes when applied to the same initial conditions and holding constant as much of the physical implementation as possible. In this work, we use a number of analytic methods to examine the CGM of the CosmoRun simulation suite (\citealp{roca-fabrega_agora_2021}, hereafter Paper III), the relevant details of which will be elucidated in Section \ref{sec:cosmorun}. This work is being developed concurrently with two additional AGORA papers also focusing on the CosmoRun simulation. The first is Roca-F\'abrega et al. (in prep), or Paper IV, which presents the final fiducial models for CosmoRun including new codes and models added since Paper III, as well as merger histories of the AGORA galaxies down to \redshift{1}. The second is Jung et al. (submitted), or Paper V, which compares the satellite populations between codes and against identical dark matter only (DMO) simulations.

While the complexity of the gas state in the CGM and dependence on so many interlocking factors make it highly unlikely that all codes will converge on the same column densities or other observational features for individual lines of sight, the carefully calibrated and specified physics and initial conditions allows profile divergences to be disentangled, or in other words to see how much each underlying variable contributes to observable quantities. This can tell us about the range of effects of modern feedback and implementation systems. For example, if significant variation takes place in metallicity distribution, this means that feedback strength and timing deliver metals from the inside to the outside of galaxies at different efficiencies. On the other hand if ion fractions are significantly different, that means that the primary effect is on cooling and heating systems causing characteristic clouds to be in a substantially different phase. We will also be looking for structure formation within the CGM, and its relationship with various ions and their kinematic distributions. 

This paper is organized as follows.
Section \ref{sec:cosmorun} describes the parameters of the codes, including initial conditions and shared physics, and gives an overview of the mechanics for each of the 8 codes participating in the study, including any existing studies of the CGM of other simulations using those codes. We also describe the analysis tools utilized in this work for creating mock observations or interpretations of the CGM. In Section \ref{sec:results} we analyze the growth and distribution of gas and metals in the CGM, including how far they spread, their usual phase, etc. We also perform analysis of observable parameters, such as absorption lines, kinematic alignment, and divergences and similarities between codes in column densities of medium-high ions. Finally, in Section \ref{sec:conclusion} we conclude the article with remarks on the essential contribution cross-code studies like this make to the field of galaxy simulations. We discuss how different codes could currently be compatible with different plausible models of the CGM, in the interest of combining their strengths to adequately resolve and populate this region in future projects.

\section{CosmoRun Simulation} \label{sec:cosmorun}

\subsection{Initial Conditions and Cosmology}

Each of the codes is designed to accept as input a common set of initial conditions (ICs), which in principle means that each of the codes should create the same zoom-in galaxy in the same location and with the same orientation. These of course will not be exactly identical, due to the stochastic elements which are built into several of the codes, but in macroscopic details they should be similar and features should be recognizable between them. The ICs are created using the software \code{music}, which uses an adaptive multi-grid Poisson solver \citep{2011MNRAS.415.2101H}\footnote{Here we use {\sc Music}'s changeset ID {\tt eb870ed}.} to create a realistic distribution of dark matter and primordial gas at a starting redshift of \redshift{100}. The zoom-in region was chosen from a large DM-only simulation such that the largest galaxy in the zoom-in region will evolve to have a virial mass of $\sim 10^{12} \,\,{\rm M}_{\odot}$ at \redshift{0}, and will not have any major merger events between the redshifts of 2 and 0.\footnote{Timing discrepancies from baryonic effects eventually led some codes to have their last major merger, supposed to take place at \redshift{2}, at around \redshift{1.9}. See Paper IV for details on timing discrepancies.} Any outside research groups, whether interested in joining as part of the Collaboration or merely to test their own code with our ICs, can freely download the \code{music} file {\it 1e12q} on the AGORA website.\footnote{See \url{http://www.AGORAsimulations.org/} or \url{http://sites.google.com/site/santacruzcomparisonproject}.} AGORA members will be happy to assist in set up and calibration of any new codes.

The cosmology used by each code is the standard $\Lambda$CDM parameters \citep{2011ApJS..192...18K_short, 2013ApJS..208...19H_short}, with an assumption of a primordial metallicity of $10^{-4} \zsun$ in each cell.\footnote{$1\, \zsun = 0.02041$ is used across all participating codes in order to follow our choice in \citetalias{kim_agora_2016} (see Section 2 of \citetalias{kim_agora_2016} for details). This has no effect on the physical conditions in \code{grackle}, which are calibrated to this value, as the total metal production by mass remains the same, though it does affect some of the plots in this work.}

Each code has a different system for refining and degrading resolution according to the local conditions, either intrinsically, as is the case for particle codes, where resolution is directly carried by particles, or by automatically refining after specific threshold requirements are met in grid codes.\footnote{\label{footnote:overdensity-of-4}Specifically, refinement takes place when an individual cell reaches a mass of four times the gas particle mass used in SPH codes (${m_{\rm gas}}=5.65\times10^4 \msun$), in order to keep grid and particle codes at roughly the same resolution, though continuity requirements for refinement does vary between codes. See Section 5.1 of \citetalias{kim_agora_2013} and section 4.3 of \citetalias{kim_agora_2016} for more details.} The resolution refinement schema for each code is listed in section \ref{sec:codes_indiv}. Overall requirements for the codes set by the ICs, however, were to have a $128^3$ root resolution in a $(60 \,\,\,{\rm comoving} \,\,h^{-1}\,{\rm Mpc})^3$ box, with five concentric regions of increasingly high resolution centered around the target halo. At the smallest, highest resolution region, it is equivalent to a unigrid resolution of $4096^3$ resolution objects, giving a minimum cell size of 163 comoving pc (around 40 physical pc at \redshift{3}). The size of this highest-resolution region is chosen to enclose all particles which will fall within $4 R_{\rm vir}$ of the target halo by $z=0$. The dark matter particles in this region are of a uniform mass (${m_{\rm DM, \,IC}} = 2.8\times10^5 \msun$), and the gas particles, for codes for that use them, have ${m_{\rm gas}}=5.65\times10^4 \msun$. For more information about this IC and other available {\it AGORA} ICs, we refer the interested readers to Section 2 of \citetalias{kim_agora_2013}, as well as Section 2 of \citetalias{roca-fabrega_agora_2021}.

\begin{table*}
\centering
\begin{tabular}{c|c c c c c c c}
Simulation & Feedback Type & Thermal Energy & Momentum & Cooling & Radiation & Stellar Mass & Stellar Mass \\
(architecture) & ~ & per SN & per SN & Delay & Pressure & \redshift{3} ($10^9 M_\odot$) & \redshift{1} ($10^9 M_\odot$) \\
\hline
\code{art-i} (AMR) & T+K, RP & $2 \times 10^{51}$ erg & $2.5 \times 10^{6} M_\odot \rm{km\,s}^{-1}\,^a$ & \textemdash & $P_{rad}^b$ & 5.0 & 17.1  \\
\code{enzo} (AMR) & T & $5 \times 10^{52}$ erg & \textemdash & \textemdash & \textemdash & 6.2 & 94.7\\
\code{ramses} (AMR) & T, DC & $4 \times 10^{51}$ erg & \textemdash & 10 Myr & \textemdash & 3.7 & \textemdash \\
\code{changa-t}$^c$ (SPH) & T & $5 \times 10^{51}$ erg & \textemdash & \textemdash & \textemdash & 16.1 & \textemdash \\
\code{gadget-3} (SPH) & T+K, RP, DC & $2 \times 10^{51}$ erg & $2 \times 10^{51}$ erg & $t_{hot}^d$ & $2.5 \times 10^{48}\textrm{M}_\odot^{-1\,e}$ & 9.2 & 35.2 \\
\code{gear} (SPH) & T, DC & $4.5 \times 10^{51}$ erg & \textemdash & 5 Myr & \textemdash & 5.9 & 38.7 \\
\code{arepo-t} (MM) & T & $2 \times 10^{52}$ erg & \textemdash & \textemdash & \textemdash & 15.1 & 65.9 \\
\code{gizmo} (MM) & T+K & $f_T \cdot 5 \times 10^{51}$ erg $^f$ & $f_K \cdot 5 \times 10^{51}$ erg $^f$ & \textemdash & \textemdash & 8.6 & \textemdash \\
\hline
\end{tabular}
\caption{Feedback style used in each code, including numerical runtime parameters when available. AMR = Adaptive Mesh Refinement, SPH = Smoothed Particle Hydrodynamics, MM = Moving Mesh, T = Thermal feedback, K = Kinetic feedback, RP = Radiation Pressure feedback, DC = Delayed Cooling feedback. The final two columns show the stellar mass within 0.15 \Rvirtext at \redshift{3} and \redshift{1}. These feedback parameters shouldn't be numerically compared to each other, and sometimes cannot, as they are not given in the same units. Still, this remains a broad overview of the breadth of implementations used in AGORA.
\\$^a$ Note that \code{art-i} is not exactly the same feedback as in \citetalias{roca-fabrega_agora_2021}, see Appendix A of Paper IV.
\\$^b$ A pressure proportional to $10^{49}\textrm{erg}\,\textrm{Myr}^{-1}\textrm{M}_\odot^{-1}$ is added to the pressure of cells containing or adjacent to cells with sufficiently high hydrogen column density and star particles younger than 5 Myr. See Section 2.2 of \cite{ceverino_radiative_2014} for details.
\\$^c$ Note that \code{changa-t} is not the same run of \code{changa} as the one in \citetalias{roca-fabrega_agora_2021}, and instead uses only thermal feedback. See Appendix B of Paper IV and section \ref{sec:changa}.
\\$^d$ See \cite{shimizu_osaka_2019} for a definition of $t_{hot}$. Generally this parameter ranges between 0.8 and 10 Myr.
\\$^e$ This value is added as heat to gas particles surrounding new star particles over a small number of timesteps, see \cite{shimizu_osaka_2019}.
\\$^f$ The fractions $f_T$ and $f_K$ are the fraction of total SN energy distributed into thermal and kinetic feedback, and depend on a number of factors according to \cite{hopkins_fire-2_2018}.
}
\label{table:feedback}
\end{table*}
\subsection{Individual codes in AGORA}
\label{sec:codes_indiv}
The codes used in this paper are summarized in depth in Papers I - IV, each paper focusing on a different aspect of how the codes work relative to different common physics implementations. \citetalias{kim_agora_2013} focuses on the details of the gravity implementation of each code, \citetalias{kim_agora_2016} focuses on the hydrodynamics and fluid dynamics solvers, and \citetalias{roca-fabrega_agora_2021} discusses the creation of stars and metals within the codes. Paper IV focuses on summarizing any changes in the active simulation setup or feedback implementation since \citetalias{roca-fabrega_agora_2021}. For convenience and to stay up to date with current developments, we also list the participating codes here, with some basics about their mechanisms and information on their most recent results, including noting any papers which focused on the CGM. 

\subsubsection{ART-I}
\label{sec:art}
The simulation code \code{art-i}, is an AMR-type grid code introduced in \cite{kravtsov_adaptive_1997}. Whenever a single cell reaches a particle or gas overdensity of 4.0 (see Footnote \ref{footnote:overdensity-of-4}), that cell splits in half along all three directions forming 8 sub-cells (codes that do this are referred to as ``octree" codes). This proceeds until the best-allowed resolution of 163 comoving pc is reached, at which point cells are no longer allowed to split. Recent work using \code{art-i} cosmological simulations includes the FIRSTLIGHT simulations \citep{ceverino_introducing_2017} with a large number of zoom-ins at high redshift. The CGM of an \code{art-i} suite was explored in significant detail in \cite{RocaFabrega19} and \cite{strawn_o_2021} for the VELA3 suite \citep{ceverino_radiative_2014,zolotov_compaction_2015}, finding that cool, inflowing streams contain mostly photoionized O~{\sc vi}, but are enclosed by Kelvin-Helmholtz interface layers \citep{mandelker_ly_2020} which contain significant quantities of collisionally ionized O~{\sc vi}. We will also point out that many of the computational and analytic tools used in this paper were first introduced in \cite{strawn_o_2021}.

\subsubsection{ENZO}
The code \code{enzo} is another AMR-type code, notable for its open-source development strategy and history \citep{bryan_enzo_2014}. It was developed alongside its native gas heating and cooling package \code{grackle} \citep{smith_grackle_2017}, which has been modified for use as a shared heating and cooling implementation used by all AGORA simulations. The most significant CGM-focused work using \code{enzo} is the development of the FOGGIE simulation \citep{peeples_figuring_2019}, as well as similar fixed-resolution halo simulations \citep{hummels_impact_2019}, which showed that resolution has very significant effects on the survival and amount of cool and cold gas found in the CGM. 

\subsubsection{RAMSES}
The \code{ramses} code is also an AMR-type octree (See Section \ref{sec:art}) code, introduced in \cite{Teyssier2002}. Current cosmological simulations which demonstrate the feedback implementation used here are shown in \cite{nunez-castineyra_cosmological_2021}, and especially \cite{augustin_emission_2019} which, focusing on the CGM of a similar \code{ramses} zoom-in simulation, found that redshift 1-2 would be a ``sweet spot'' for observations of the CGM in emission with new telescopes now coming online.

\subsubsection{CHANGA-T}
\label{sec:changa}
\code{changa} is a particle SPH code, where fluid interactions are mediated between multiple ``smoothed particles." It is is a redevelopment of the code \code{gasoline} \citep{menon_adaptive_2014,Wadsley2017} with a different architecture. This code has been recently used for the ROMULUS simulation series, summarized in \citep{jung_massive_2022}. The CGM of several ROMULUS halos was recently analyzed and categorized a large number of different phases and dynamic modes in \cite{saeedzadeh_cool_2023}. 

We have changed the name to \code{changa-t} to indicate a different version from the one used in \citetalias{roca-fabrega_agora_2021}. In that paper, we ran a version of \code{changa} with so-called ``superbubbles," a form of feedback that superheats small regions near supernovas \citep[see][]{keller_superbubble_2014}, while the version shown here has only thermal feedback, as visible in Table \ref{table:feedback}. Both versions of \code{changa} were run with the CosmoRun ICs, with a comparison between the two shown in Appendix B of Paper IV. We focus on this version here because it was more easily accessible at the time of submission of this paper and could be analyzed more straightforwardly, however further comparison between the CGM of the two versions would be an interesting topic for future work.

\subsubsection{GADGET-3}
The next SPH-type code is \code{gadget-3}, a highly versatile code with many different offshoots, with gravity computed by the tree-particle-mesh method. \code{gadget3-osaka}, referred to in this paper as \code{gadget-3} \citep{aoyama_galaxy_2017, shimizu_osaka_2019} is one of several offshoots of the SPH code \code{gadget} (Generations 1 and 2 were showcased in \citealp{springel_gadget_2001} and \citealp{springel_cosmological_2005}, respectively). The code used in this paper uses the feedback system adapted from \cite{shimizu_osaka_2019}. 

Previous studies of the CGM in \code{gadget-3} include \cite{oppenheimer_bimodality_2016}, which analyzed the EAGLE simulation and found that in their codes, O~{\sc vi} was not necessarily connected to galaxy star formation as inferred from \cite{tumlinson_large_2011}. \citet{nagamine_2021} also studied the distribution of neutral hydrogen in the CGM, and showed that varying treatment of feedback can cause about 30\% variations in the Ly$\alpha$ flux decrement around galaxies. 

\code{gadget-4} \citep{springel_gadget-4_2022} is also in current use 
\citep[e.g.,][]{romano_2022a,romano_2022b}, and has expressed interest in pursuing the AGORA project. It will be included in future papers after completion of the rigorous calibration required by CosmoRun.

\subsubsection{GEAR}
The code \code{gear} \citep{revaz_dynamical_2012} is another SPH code. While originally based on \code{gadget-2}, it contains a number of improvements and possess its own physical model \citep{revaz2016,revaz_pushing_2018}. \code{gear} uses the improved SPH formulation of \citet{hopkins_general_2013} and operates with individual and adaptive time steps as described in \citet{durier_implementation_2012}. Star formation is modelled using a modified version of the stochastic prescription proposed by \citet{Katz1992} and \citet{Katz1996}, where stars form in unresolved regions, and which reproduces the \citet{Schmidt1959} law. Stellar feedback includes core collapse and type Ia supernovae \citep{revaz2016}, where energy and synthesised elements are injected into the surrounding gas particles using weights provided by the SPH kernel. To avoid instantaneous radiation of the injected energy, the delayed cooling method is used \citep{2006MNRAS.373.1074S}. The released chemical elements are further mixed in the ISM using the smooth metallicity scheme \citep{Wiersma2009enrichment}. 

The \code{gear} physical model has been mainly calibrated to reproduce Local Group dwarf galaxies \citep{revaz_pushing_2018,Harvey2018,Hausammann2019,Sanati2020} and ultra-faint dwarfs \citep{Sanati2023} and in particular their chemical content.

\subsubsection{AREPO-T}\label{sec:arepo}
The \code{arepo} code operates using an unstructured moving mesh, which is generated dynamically according to density and velocity, allowing it to evolve resolution naturally while still solving Euler equations on cell faces as in grid codes \citep{springel_e_2010}. Major recent \code{arepo} projects include Illustris-TNG \citep{pillepich_simulating_2018} and Auriga \citep{grand_auriga_2017}. Analysis of the CGM of the former was given in \cite{nelson_resolving_2020}, finding that magnetic fields could be essential to cold clouds surviving in the halo, and of the latter in \cite{van_de_voort_effect_2021}, which found in a zoom-in simulation that resolution was essential to resolving cold and cool neutral gas in the CGM. 

Like \code{changa-t} (Section \ref{sec:changa}), we have adopted the name \code{arepo-t} in this paper to indicate this run uses only thermal feedback. Another version with a different feedback system has also been run on the same initial conditions by the Collaboration. That run contains a more complex schema for stellar wind propagation \citep[see Section 2.3.2 of][]{pillepich_simulating_2018} and is compared to the version here in Paper IV, Appendix B. In this work, we focus on the thermal-only version because it was somewhat faster to calibrate and simpler to analyze, making it more accessible at the time of publication of this paper. Direct comparison between the CGM of \code{arepo}'s thermal and IllustrisTNG-like wind models will be considered as a future project by the AGORA collaboration.

\subsubsection{GIZMO}
Finally, \code{gizmo} is a mesh-free code based on a volume partition scheme, in which particles represent cells with smoothed boundaries. Despite being a descendant of \code{gadget-3}, \code{gizmo} is somewhat similar in spirit to \code{arepo}, where the Euler equations are solved as in grid codes across effective faces shared between nearby particles. The actual scheme employed in the \code{gizmo} runs for this comparison is the finite-mass one, in which cells are not allowed to exchange mass through the faces. 

The Simba \citep{dave_simba_2019} and FIRE-2 \citep{hopkins_fire-2_2018} projects are examples of high-resolution zoom-in \code{gizmo} simulations. These works found that in the CGM, cool inflows generally reached temperature equilibrium quickly and are not very sensitive to the heating implementation, while hotter gas has a cooling time longer than the dynamical time and therefore its state depends more sensitively on this implementation.

\subsection{Common, Code-independent Physics} \label{sec:common-phy}

Much of the physics in the operation of the codes is fixed, and each aspect of this was thoroughly calibrated in the process described in \citetalias{roca-fabrega_agora_2021}.\footnote{Note that some CosmoRun models, specifically \code{art}, \code{changa-t}, and \code{arepo-t}, were either not present in \citetalias{roca-fabrega_agora_2021}, or are different than the ones used in that work. Calibration details for for the codes shown in this work, and full descriptions of their star formation and feedback systems, are instead given in Appendices A and B of Paper IV.} While hydrodynamic and gravitational solvers are intrinsically tied to individual codes, gas heating and cooling parameters are fixed by the common package \code{grackle}\footnote{Version 3.1.1} \citep{smith_grackle_2017}, and the details of the \code{grackle} runtime parameters were shown in Section 3.1 and the process of calibration with each code was shown in Section 5.2 (Figures 4 and 5) of \citetalias{roca-fabrega_agora_2021}.

A pressure floor requires the local Jeans length to be resolved at all times, in order to prevent unphysical collapse and fragmentation, and each code was given a minimum cell size (for AMR codes) or gravitational softening length (for SPH codes). More details on these conditions can be found in \citetalias{roca-fabrega_agora_2021}, specifically sections 3.1 and 4. In this paper which focuses on the much lower-resolution CGM region, we are interested in not just the highest available resolution, but also the specific pattern of the resolution degrading as the simulation moves away from the galaxy center. 

\begin{figure*}
  \includegraphics[clip,width=\linewidth]{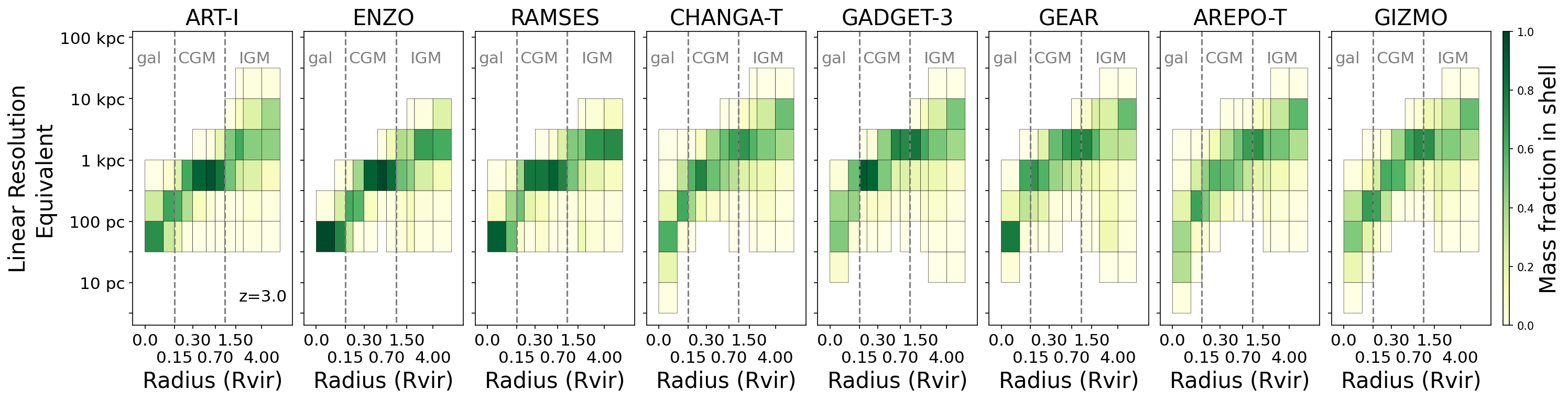}
\caption{Resolution of all 8 AGORA codes at \redshift{3}. In each shell of increasing size, color shows the mass fraction contained in ``linear resolution equivalent'' bins of width 0.5 dex, normalized within columns. For grid and moving mesh codes, ``linear resolution equivalent'' is defined as cell volume raised to the 1/3 power. For particle-type codes, it is instead defined as ``effective volume'' (particle mass divided by particle density) to the 1/3 power. See Section \ref{sec:common-phy} for more details.}
\label{fig:resolution}
\end{figure*}

In Figure \ref{fig:resolution} we show the increase in the effective size of resolution elements as a function of distance to the galaxy center for each code. All codes were found to show a general degradation in resolution with distance, and mostly convergent with one another. Generally, all codes have a resolution of between 30--300 pc within 0.15 \Rvirtext (considered to roughly represent the ``galaxy''), between 100 pc -- 3 kpc within 1.0 \Rvirtext (representing the ``CGM''), and between 300 pc -- 10 kpc outside 1.0 \Rvirtext (the ``IGM,''), with the outer boundary of the IGM taken to be at 4.0 \Rvirtext in order to stay within the Lagrangian region defined in the ICs. A few resolution differences between the codes persist, however, mostly as a result of their general hydrodynamical mechanism. SPH codes are not as strongly constrained by either resolution ceilings or floors, because the free motion of particles is paramount. While particle masses are chosen in order to force a certain mass resolution, if gas particles cluster together into a small region, they will effectively resolve that volume at a better resolution than the best-allowed volume resolution for AMR codes, and thus can be more detailed within the internal galaxy structure.\footnote{However, SPH code gravity is still limited by the smoothing size of particles, which is constrained to be greater than or equal to to the best resolution of grid codes -- ``effective volumes'' smaller than this size are not fully self-consistent.} The disadvantage of this free motion is that in low density regions such as the CGM the effective resolution in particle codes is worse than in grid codes, which have their resolution-degradation suppressed by the strict requirements for cell recombination. Moving mesh codes remain somewhere in between these two outcomes. Within the IGM, all types of codes have very similar outcomes. 

All codes are given the same requirements to form stars, though how those requirements are implemented can vary greatly. The code groups are each asked to determine, according to their code's design and particle generation format, the stochastic or deterministic nature of this process. This takes place at a threshold number density of $1 \textrm{cm}^{-3}$. The mass of each star particle formed also determined by the individual processes, only requiring a minimum mass of $6.1\times 10^{4} \textrm{M}_\odot$. Details on the requirements for star formation within the codes in CosmoRun are given in \citetalias{roca-fabrega_agora_2021}.

Unlike in \citetalias{kim_agora_2016}, where the form of stellar feedback was specified in an idealized galaxy disk, in the CosmoRun simulation of Papers III--VI (this work), we allow each supernova's schema for injection of metals, mass, and energy into the nearby gas to be as close as possible to the version most commonly used by that code group in comparable simulations. We do require some top-level parameters to be the same. Specifically, we require each supernova event to release at least $10^{51}$ ergs of thermal energy, 14.8 $\msun$ of gas, and 2.6 $\msun$ of metals. This change was detailed further in \citetalias{roca-fabrega_agora_2021}. Different codes add many different effects or implement feedback in different ways, as shown in Table \ref{table:feedback}. 

Notably, we use the ``thermal-only'' models analyzed in Paper IV Appendix B for \code{changa} and \code{arepo}. In addition to the logistical reasons stated in sections \ref{sec:changa} and \ref{sec:arepo}, this is useful because it allows us to examine one example of the CGM that results from each code architecture using simple thermal-only feedback, these being \code{enzo} (AMR), \code{changa-t} (SPH), and \code{arepo-t} (MM). 

\subsection{Shared Analysis Tools}
\label{sed:analysis_tools}
The most important analysis tool for this work is the highly versatile simulation analysis code \code{yt}. This code was first developed in \cite{turk_multi-code_2011}, and significant improvements to \code{yt} were integrated by AGORA collaborators during the process of writing Papers I, II, and III, alongside many others. The code has reached widespread adoption in the cosmological simulation community, and engagement from that community has led to significant improvements in all aspects of the code. The most significant update since \citetalias{roca-fabrega_agora_2021} to \code{yt} is the ``demeshening", where particle codes were integrated much more naturally into the architecture, which was designed primarily for use on grid codes (Turk et al., in prep). We also rely heavily on a \code{yt}-based CGM tool \code{trident} \citep{hummels_trident_2016}, which makes sightline generation significantly easier, implements ion fractions using a lookup table from the photoionization code \code{cloudy} \citep{ferland_2013_2013,ferland_2017_2017}, and has efficient functions for both generating and analyzing realistic spectra.

With these two programs powering our back-end analysis, we have developed a user-oriented frontend tool \code{agora\_analysis},\footnote{\url{https://github.com/claytonstrawn/agora\_analysis}, using the version released as \cite{strawn_claytonstrawnagora_analysis_2023}.} which is integrated into the shared supercomputer architecture\footnote{Simulations are currently stored for analysis on the US Department of Energy (\href{https://nersc.gov/}{NERSC}) supercomputer, and will be released publicly following publication of Paper IV.} to make accessing each simulation snapshot and any necessary metadata for that snapshot (center coordinates, \Rvirtext, bulk velocity vector, angular momentum vector) very straightforward for use by any collaborators or interested parties. \code{agora\_analysis} also includes scripts for creating most of the images in this text, besides the ones which use individual sightline data for which there is another package \code{quasarscan}. As an important point here, by default \code{agora\_analysis} will calculate the sizes of different regions using a virial radius which is the average of all eight codes' individual virial radii generated using \code{rockstar} \citep{behroozi_rockstar_2013}. At a fixed stellar mass and with a fixed environment, it was decided that to include significantly more (up to $\sim 1.5$ times, at most) volume in some codes would detract from the comparison, especially when considering the number of satellites of the main halo (Paper V). So, all virial radii and derived quantities taken within the 0.15 \Rvirtext\, edge of the galaxy or the 1.0 \Rvirtext\, edge of the CGM are shared among all 8 codes, even though individual virial radii have been calculated for each.

Finally, \code{quasarscan}\footnote{\url{https://github.com/claytonstrawn/quasarscan}, using the version released as \cite{strawn_claytonstrawnquasarscan_2023}.} is a random sightline generator and analysis tool, first introduced in \cite{strawn_o_2021}. It creates approximately 400 sightlines through the CGM by placing sightline start points on an enclosing large sphere ($\sim 6.0\Rvir$) at a discrete set of polar and azimuthal coordinates, and the vector from the galaxy center to the start point is normal to a ``midpoint'' plane within which the distance to the galaxy center will be equal to the sightline impact parameter. A midpoint is then selected from that plane at one of a discrete set of impact parameters from 0 to 1.5 \Rvirtext. The probability of each impact parameter and polar angle is weighted so that the lines comprehensively sample the area within that radius, i.e. higher impact parameters are more likely. The sightline is then projected from the starting point through that midpoint, and ends back on the aforementioned large sphere, on the opposite side of the halo.

Each line of sight integrates a set of ions of interest (here, Si~{\sc iv}, C~{\sc iv}, O~{\sc vi}, and Ne~{\sc viii}) to calculate a column density. Furthermore, we calculate the overall metallicity, and the mean and peak densities along the line. For a small subset of sightlines, physical spectra are also projected and saved, for analysis with \code{trident}'s built-in Voigt profile fitter (Figures \ref{fig:spectra_z3}, \ref{fig:spectra_z1}, and \ref{fig:spectrum_analytics}). 

\begin{figure*}
  \includegraphics[clip,width=\linewidth]{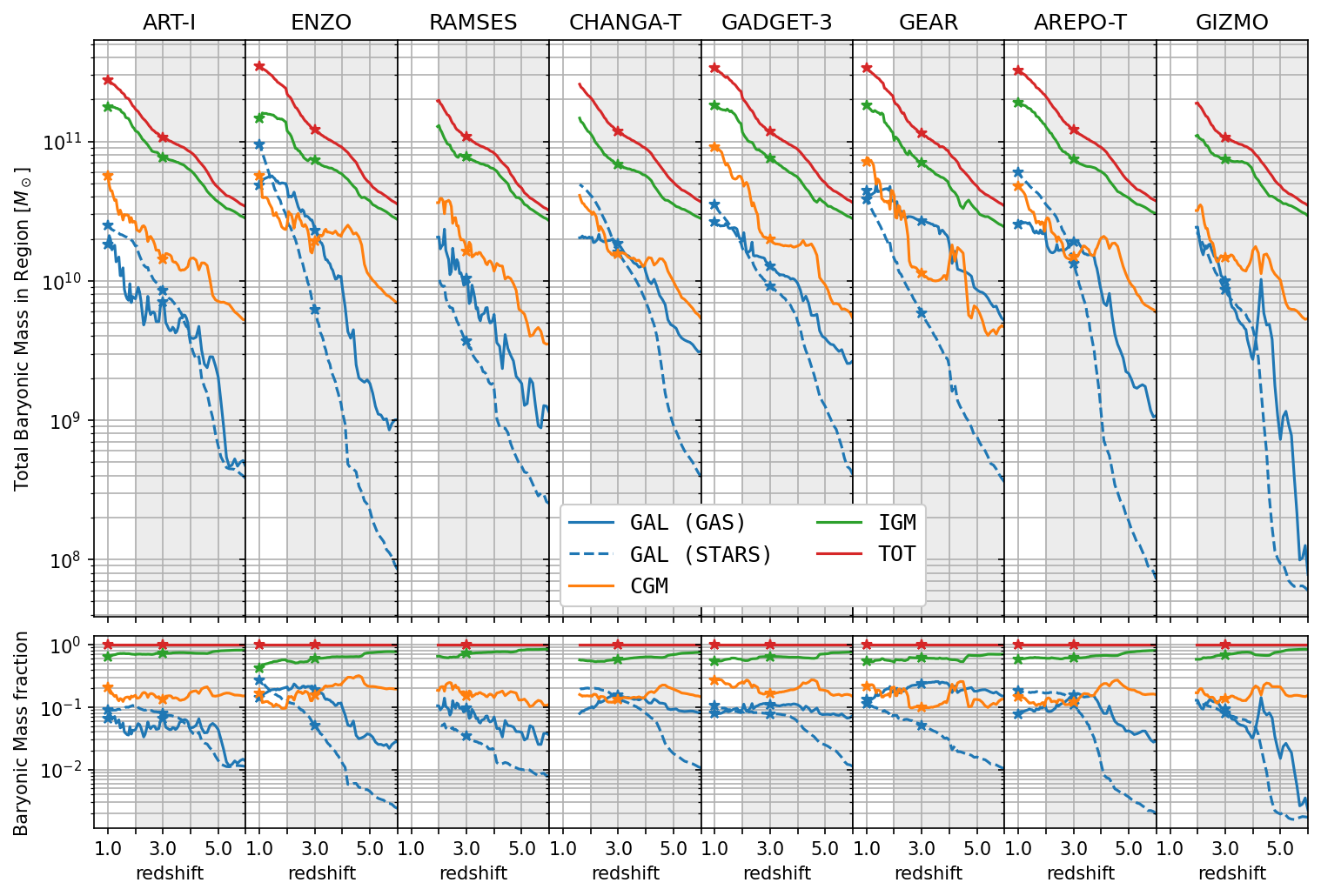}
\caption{Current distribution with redshift (evolving from right to left) of gas mass in the galaxy, CGM, and IGM, both in solar masses, top, and as a fraction of the total, bottom. ``GAL" (galaxy) refers to the region from 0.0 - 0.15 $\Rvir$, ``CGM" refers to 0.15 - 1.0 $\Rvir$, and ``IGM" is defined as the region between 1.0 - 4.0 $\Rvir$. Starred points are added to each line at redshifts 3 and 1, to guide the eye when comparing to other plots in this work. Additionally, the shaded region down to \redshift{2} is shaded to indicate the epoch reached by all 8 codes. Inside the galaxy, the total mass is split between stars and gas.}
\label{fig:gas_mass_dist}
\end{figure*}

\section{Results of CGM study}
\label{sec:results}

Because of how sensitive the CGM's observables are to so many different variables, it is worth reiterating the design philosophy of the AGORA project. In \citetalias{kim_agora_2016} we have already established that the implementation differences between codes in idealized conditions are minimal. So, any significant differences between codes are likely to be a result of their different choices of stellar and supernova feedback at least as much as their underlying hydrodynamical and gravity solver. For the convenience of the reader, we will continue to refer to different codes by code name, rather than by referring to the feedback mechanism explicitly, except where the feedback appears to have clear effects on the outcome. This means that other simulation groups using a code in AGORA with a different feedback implementation are cautioned to be careful when comparing their simulation to the CosmoRun results for their code. As mentioned above, these are the initial feedback models, and several codes have already run the same ICs with new feedback prescriptions, which will be added to the AGORA public data release and will be analyzed in future works. We will also comment that at the level of detail of individual particles, streams, or other features, there are inherent stochastic and numerical effects, which means that some details might not be the same even between runs of the same code. This means, however, that we should be careful interpreting very specific objects, such as lines of sight, slices, and projections, and that instead, averages, profiles, and phase diagrams will be more robust to stochastic effects.

\subsection{Differences in metal distribution and gas state}
\label{sec:cgm-metals}
The most striking feature of the different codes for their observable CGM is precisely the difference in mass and metal distribution out to \Rvirtext and beyond, which depends strongly on feedback mechanism and code architecture. In Figure \ref{fig:gas_mass_dist}, we show the evolution of the gas mass distribution throughout all eight models over time, both as raw masses (top) and as a fraction of the total (bottom). The four components of gas mass are as follows:
\begin{enumerate}
\item \texttt{"GAL (GAS)"}: gas within 0.15 \Rvirtext
\item \texttt{"GAL (STARS)"}: stellar mass within 0.15 \Rvirtext
\item \texttt{"CGM"}: gas (and stars) between 0.15 and 1.0 \Rvirtext, however only a small number of star particles are present
\item \texttt{"IGM"}: gas (and stars) between 1.0 and 4.0 \Rvirtext, however as with the previous item, only a very small number of star particles are present. 
\item \texttt{"TOT"}: Total gas and star mass in the entire 4.0 \Rvirtext\, enclosing sphere.
\end{enumerate}
We can notice here that all eight codes agree remarkably well in the total gas (red curve) at both redshifts \redshift{3} and \redshift{1}. Note that not all codes reach redshift \redshift{1}, meaning that the codes do not necessarily agree at their own ``last'' points.\footnote{Different codes reach different final times not based on their performance or efficiency, but rather because the supercomputing resources available for each code group varied.} All of the AGORA galaxies are dominated by gas in the IGM throughout cosmic time, as expected due to it containing more than 98 percent of the total analyzed volume, and due to primordial gas which continues to inflow along cosmic filaments into the ``IGM'' region. Within the galaxies, there is significant variation between retaining more mass in stars or gas over time, with stellar mass eventually eclipsing gas mass in \code{art-i},  \code{enzo}, \code{changa-t}, \code{gadget-3}, and  \code{arepo-t}. Interestingly, the CGM mass (orange) remains more consistent among codes, even though whether the CGM is overall larger or smaller than the galaxy mass is not. The CGM contains in some codes more mass than galactic stars and gas combined, even being overtaken by stars alone in \code{enzo}, \code{changa-t}, and \code{arepo-t}, notably the three codes using thermal-only feedback. Additionally, notice that all codes have an extremely ``bursty'' accretion pattern into the CGM with the mergers that take place at \redshift{5} and less noticeably at redshift $z\sim 2$.

\begin{figure*}

\includegraphics[clip,width=\linewidth]{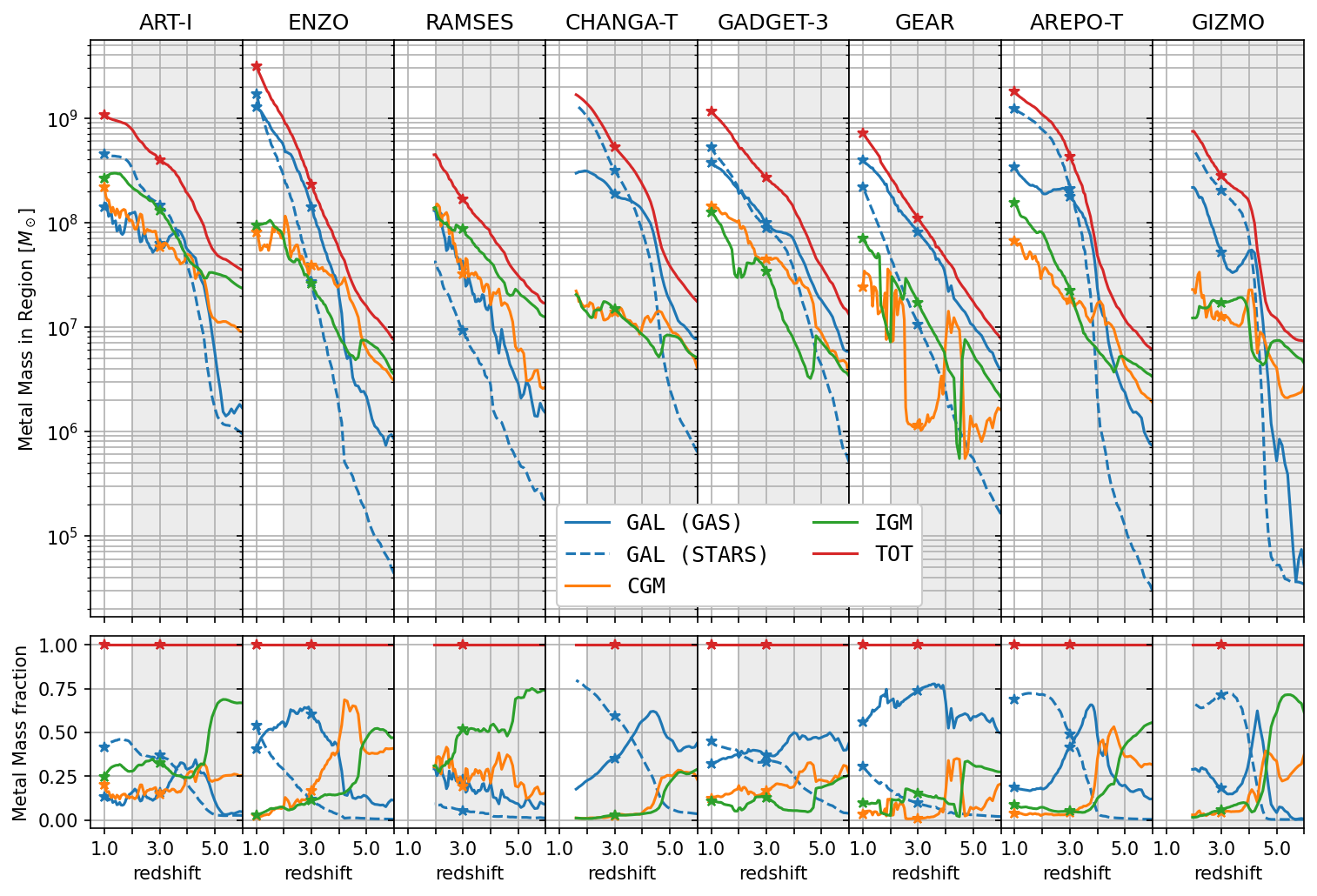}

\caption{Like Figure \ref{fig:gas_mass_dist}, but now tracing the total mass of metals in and around the main AGORA galaxy in each simulation.}
\label{fig:metal_mass_dist}

\end{figure*}

In Figure \ref{fig:metal_mass_dist} we examine the distribution and evolution of metals in the different regions with time. As in Section \ref{sec:common-phy} we have selected to take 4.0 \Rvirtext as the outer boundary of the IGM because inclusion of any regions outside this distance creates unphysical metal distribution results. This arises because with integration of large volumes, the metallicity floor for the AMR codes results in substantial metals far from any meaningful sources, while the total in SPH codes is much lower. Within this sphere, all metal mass can be assumed to originate in local stars, either within the central galaxy or in satellites. 

Overall, the total metal creation (red) is relatively consistent, though not as consistent as we would expect, given the requirements on each code and the closeness of their star formation rates. The effective metal yields are given in Table 1 of \citetalias{roca-fabrega_agora_2021}, and generally the yield is a metal mass of 0.033 $\msun$ for each 1.0 $\msun$ of stellar mass. The exception to this is \code{gear}, where the metal production (yield 0.015) could not be detached from the star formation prescription. This means that metal production in \code{gear} is consistently at least a factor of two below the other codes, though it is worth noting that it is more than a factor of two below at other times, indicating it is not only the yield which suppresses metal production. At redshifts approaching \redshift{1}, \code{art-i} slows this production significantly. This is likely due to an oncoming quenching period where star formation slows down in most codes, and which will be a topic of future AGORA papers.

Total metal production is within a factor of 4 at redshift \redshift{2} (between \code{ramses} and \code{changa-t}), and retains about the same range at \redshift{1}, but now between \code{gear} and \code{enzo}. Overall, \code{enzo}'s consistently high star formation causes a dramatic turnaround from the slow start; in \citetalias{roca-fabrega_agora_2021} it was noted that \code{enzo} had the \textit{lowest} stellar mass of all eight codes in CosmoRun at \redshift{4}. Here it has the highest SFR by a decent margin, with only \code{changa-t} coming close, and already slowing down by \redshift{1.5} (Figure \ref{fig:gas_mass_dist}). This has a complex relationship with \code{enzo} having the strongest purely thermal feedback of all AGORA codes, which clearly suppresses star formation at early times but which then allows additional star formation at later times. Further discussion of the star formation rates as a function of time and feedback process can be found in Paper IV.

Interestingly, there is no consistent pattern as to whether most metals within the galaxy remain locked into stars, effectively inaccessible to any kind of gas mixing (\code{art-i}, \code{changa-t}, \code{arepo-t}, \code{gizmo}), or whether most metals are in the ISM and thus could be subject to outflows and/or recycling (\code{ramses} and \code{gear}, codes both using T, DC feedback), while \code{enzo} and \code{gadget-3} keep the ISM and stellar metal mass roughly equal.  

Another striking feature of this plot is how some codes, in particular \code{art-i} and \code{ramses}, send comparable amounts of metals into the IGM or CGM as remain inside the galaxy, including star contributions, while most codes keep the vast majority inside the galaxy. With regard to \textit{how far} the average metals go, we can note that regardless of how much of the metal mass leaves the central galaxy, generally metals that do leave become roughly equally divided between the IGM and CGM, with the exception being the fast-outflowing \code{art-i} and \code{arepo-t} galaxies which eject metals so quickly from the ISM that they flow through the CGM and immediately leave, leading the IGM to dominate the metal distribution, though as \code{art-i} approaches \redshift{1}, the metals slow down and seem to return to the CGM. Metal diffusion and transportation processes depend in complex ways on code architectures, as discussed in detail in Section 3.2 of \citetalias{roca-fabrega_agora_2021}. In grid codes, diffusion over surfaces is built in with solving the Reimann problem on each cell interface, while in particle codes, diffusion is often implicit in the smoothing procedure. Moving mesh codes can provide either explicit or implicit diffusion depending on their architecture, see \citetalias{roca-fabrega_agora_2021} for explanation of \code{gizmo} and Paper IV, Appendix B for one of \code{arepo}.

\begin{figure*}
    \includegraphics[clip,width=\linewidth]{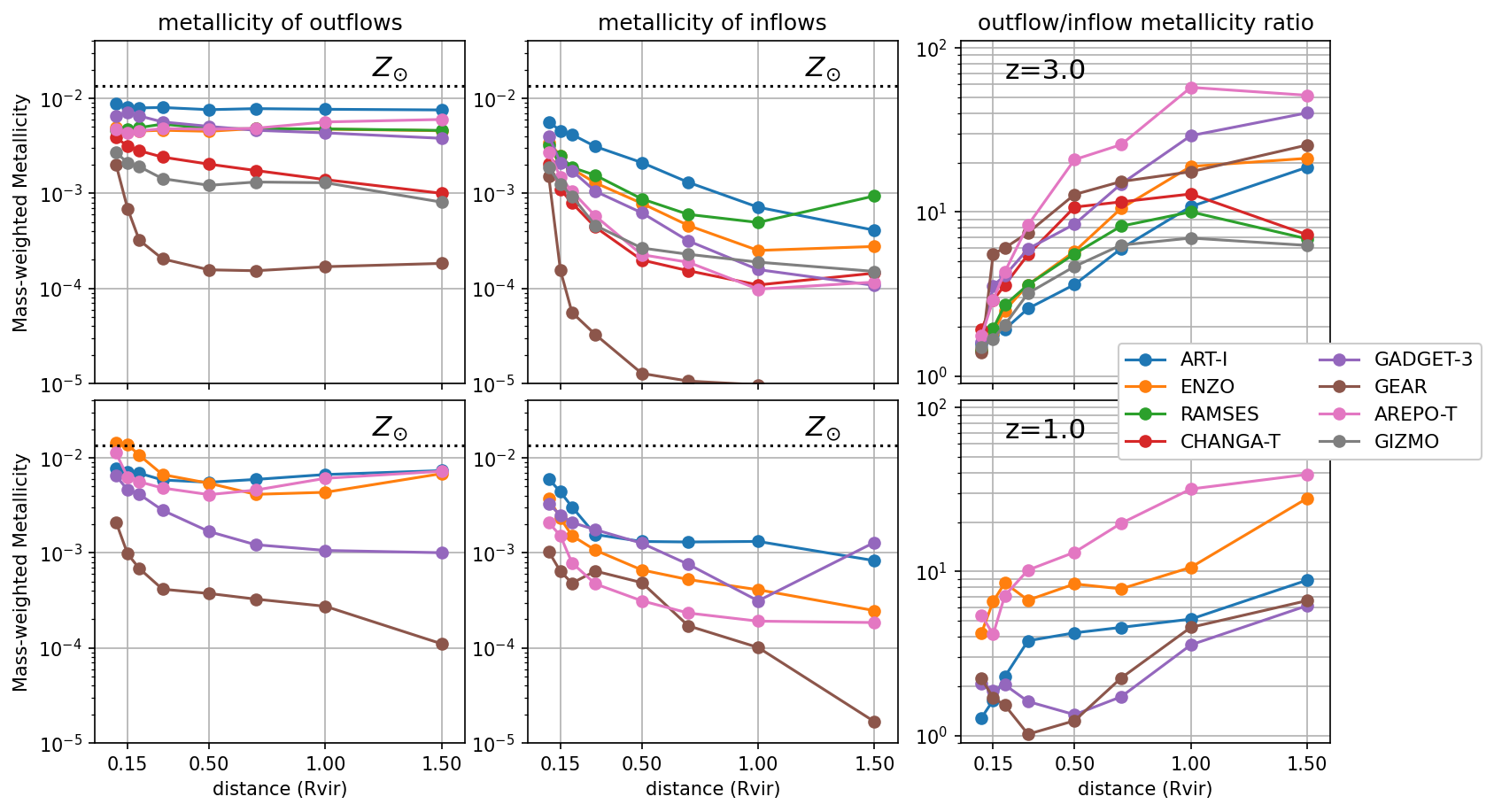}
\caption{Here we show the metallicities of outflowing (left) and inflowing (center) gas elements (cells or particles, depending on the code architecture), as a function of radius. Top row is redshift $z=3$, bottom row is redshift $z=1$. Right: outflow metallicity divided by inflow metallicity with radius. This is much more similar between codes than the individual metallicities of the two phases.}
\label{fig:metallicity_inflow_outflow}
\end{figure*}

Finally, we will analyze a property which is common to all eight codes. Namely, in Figure \ref{fig:metallicity_inflow_outflow} we show the overall metallicities of the outflowing and inflowing gas elements (cells or particles) in the left and center columns. In the outflowing gas column at \redshift{3}, while there is an approximately two orders of magnitude difference between the highest and lowest metallicities, all codes remain approximately flat with radius outside the galaxy, declining only by a factor of 3 at most (in \code{changa-t}). \code{gear} remains constant outside of 0.5 \Rvirtext, but declines significantly within that region, indicating that with the feedback process implemented in that code, only a small amount of ejected gas reaches the virial radius (in addition to the previously mentioned factor of 2 lower yield, see \citetalias{roca-fabrega_agora_2021}). Inflowing gas has signficantly lower metallicities overall in all codes, with a significantly stronger decline with radius. In the third column of Figure \ref{fig:metallicity_inflow_outflow} we show that the ratio of inflowing to outflowing metallicity is much more closely constrained, with less than an order of magnitude difference between the codes. Figure \ref{fig:metallicity_inflow_outflow} suggests that all codes have outflows and inflows interacting with similar dynamics, which causes inflows to significantly increase in metallicity as they approach the central galaxy. The similarity between codes on the ratio of outflows to inflows, combined with the very different total metallicity of each, suggests that it is indeed the feedback systems, rather than the overall code architecture (which would control inflow-outflow dynamics) which affect the distribution of metals. Previously, it was found that, in some cosmological simulations \citep{mandelker_ly_2020, strawn_o_2021}, cool inflows entrained metals from the hot outflowing material, so that when they fed the galaxy they were barely more metal-poor than the hot outflows, leading newly formed stars and cool gas to be generally not ``pristine." These results suggest something broadly similar here, and so gas entering the galaxy from outside is likely to be only mildly more metal-poor than the ISM itself.

\subsection{Comparison Snapshot Analysis}
\label{sec:comparison}
\begin{SidewaysFigure}

\includegraphics[clip,width=\linewidth]{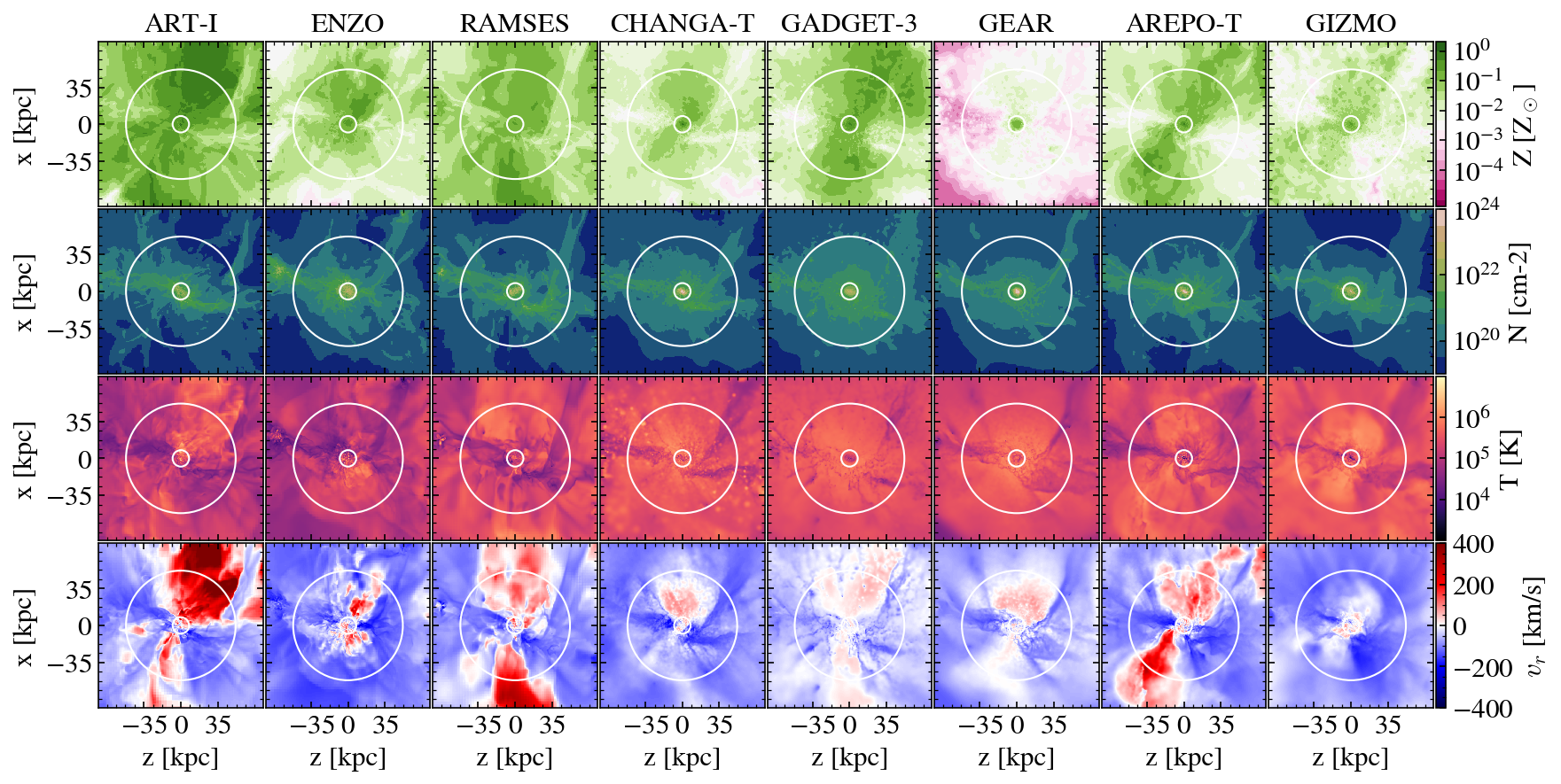}
\caption{Mass-weighted Projection Plots at redshift \redshift{3} of all eight codes in four fields, out to 1.5 times the average virial radius of all codes (\Rvirtext = 53 kpc). Inner and outer white circles represent 0.15 and 1.0 \Rvirtext, respectively. Rows (from top) are metallicity, column density, temperature, and radial velocity $v_r$, where $v_r > 0$ represents outflows and $v_r < 0$ represents inflows. Projections are aligned with simulation box axes, rather than angular momentum (face-on vs edge-on) for global consistency. Cool, dense inflows are visible along the left-right axis in each code, and metal-rich outflows along the up-down axis.}
\label{fig:projection_8codes_z3}
\end{SidewaysFigure}

\begin{figure*}
    \centering
  \includegraphics[clip,width=0.8\linewidth]{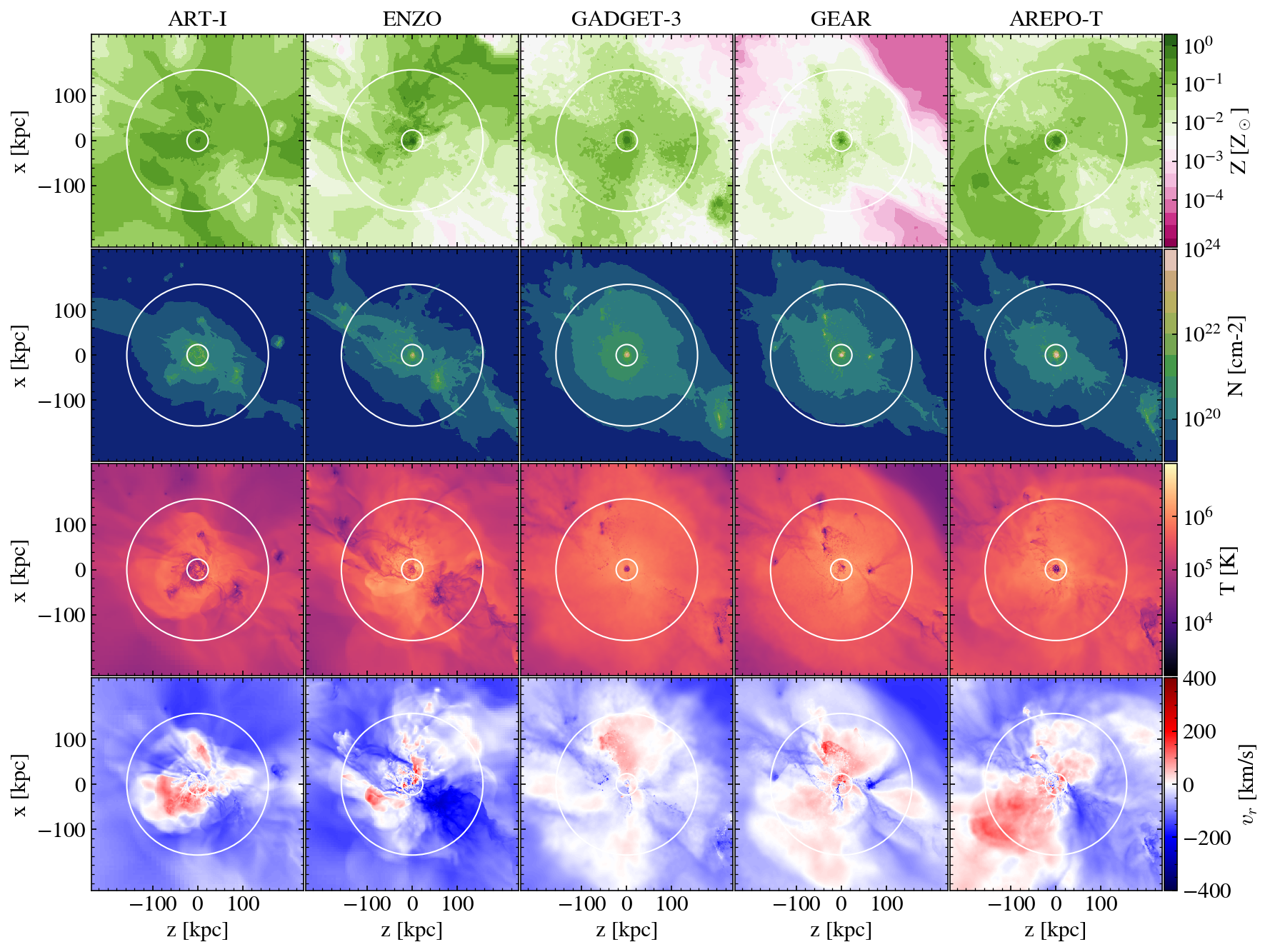}
\caption{Identical to Figure \ref{fig:projection_8codes_z3}, but for five codes at redshift $z=1$. At this redshift, \Rvirtext = 153 kpc.}
\label{fig:projection_5codes_z1}
\end{figure*}

Here we will perform a detailed analysis of a single snapshot for eight codes at redshift \redshift{3}, and five codes at redshift \redshift{1}. These redshifts are chosen to avoid any effects from the timing discrepancies of mergers at redshifts 4 and 2 (See Paper IV for details on the timing discrepancies). First, we analyze a projection plot at a particular viewing angle for Figures \ref{fig:projection_8codes_z3} and \ref{fig:projection_5codes_z1}. The rows of this plot are metallicity, column density, temperature, and radial velocity, respectively, with columns representing different codes. Note that these plots are chosen to be axis-aligned to show shared structural features. Face-on and edge-on figures are available in Paper IV. We also elected to use thin mass-weighted projections rather than slices to facilitate straightforward comparisons, which due to stochasticity, timing discrepancies and minor numerical effects, have features which are rarely aligned into identical planes, even if they are largely the same. A good example of both is the cool streams which are visible in the temperature projection (third row from top) in each code, which are clearly relatively similar between codes here; with slightly different image parameters these streams would only appear in some panels. As noted in \cite{stewart_high_2017}, in many simulation codes, including several codes showcased here, angular momentum is primarily built up through these inspiraling flows which are connected to the cosmic web.

At \redshift{3}, there are many similarities between the snapshots. The mass structure is broadly the same in each code, with the main star formation fuel \textemdash cool, dense, inflowing streams \textemdash being approximately $z$-axis-aligned, with $N\sim10^{21} \textrm{cm}^{-2}$, and a hot outflowing bulk medium elsewhere. Average column density in the galaxy region is at $10^{23} \textrm{cm}^{-2}$ and above in all codes except \code{art-i}. Within the CGM, average densities outside of the streams are around $N\sim10^{20} \textrm{cm}^{-2}$, with only \code{gadget-3} seeming to have a significant filling out to the virial radius with higher density. In temperature, there is a fairly substantial difference between the grid and particle codes, with significantly more cool gas visible in \code{art-i}, \code{enzo}, \code{ramses} and \code{arepo-t}. \code{ramses} and \code{arepo-t} have particularly strong contrasts, containing cooler high-density clouds and a hotter low-density bulk. Moving mesh codes have behavior somewhat in between the two styles, with \code{gizmo} more closely resembling the particle codes and \code{arepo-t} more closely resembling grid codes.

In this axis-aligned image some important differences can be very subtle, such as that in some codes (\code{art-i}, \code{enzo}, \code{ramses}, \code{arepo-t}), the inflowing stream from the center right merges with the other stream on the top right near the virial radius and gives the impression of a single stream entering the halo, while in others  (\code{changa-t}, \code{gadget-3}, \code{gizmo}), the three streams generally merge much closer to the galaxy, appearing as more or less separate valves for inflow. This is a highly stochastic effect, which depends sensitively on the plane chosen and timestep. The most significant difference is in fact the volume and metallicity of the outflow structure. An extremely visible effect which distinguishes particle codes from grid codes is how fast gas is ejected, as seen in the radial velocity images in Figure \ref{fig:projection_8codes_z3} (bottom row). While a biconical outflow structure is visible in all codes (though very faintly in \code{gizmo}), the difference between the extremely fast speeds in the grid codes and the much slower speeds in the particle codes leads to metals being more uniformly distributed in grid codes out to large distances, as also noted in \cite{shin_how_2021}. Examination of larger-scale plots, shown in Appendix \ref{sec:appendix}, demonstrates that the maximum spatial extent of metals in these codes is a sphere of about 4.0 \Rvirtext.

\code{art-i} and \code{ramses} are by far the strongest, and send gas sometimes with supersolar metallicities at speeds on the order of 100s of km/s, with \code{arepo-t} containing similar speeds in somewhat narrower outflow jets, (note that Figure \ref{fig:projection_8codes_z3} is a mass-weighted projection, so these values are significantly diluted by slow-moving or slowly infalling gas along the projection lines-of-sight). The \code{gadget} and \code{changa-t} snapshots have similarly shaped high-metallicity biconical outflows, but much slower, and \code{gizmo} has even weaker outflows. While \code{enzo's} outflowing gas is as fast as the other grid codes, its much narrower structure means fewer overall metals leave the virial radius. Finally, \code{gear} has significantly less metals sent into the CGM than any of the other codes, due to the low yield, highly concentrated center and relatively slow outflows. 

By \redshift{1}, (Figure \ref{fig:projection_5codes_z1}) a number of changes have taken place. The higher density gas filling the virial radius, seen before in \code{gadget-3} has also happened in \code{gear}. The grid codes, here including \code{arepo-t}, remain largely filamentary, most visible in low-metallicity in the top row. Grid codes retain both faster inflows and outflows, and over the time from \redshift{3} to \redshift{1}, we can see that both particle codes have significant metallicities only about out to the virial radius (with \code{gear} somewhat less than \code{gadget-3}), while the grid codes have effectively filled the visible IGM. 

\begin{figure*}
  \includegraphics[clip,width=\linewidth]{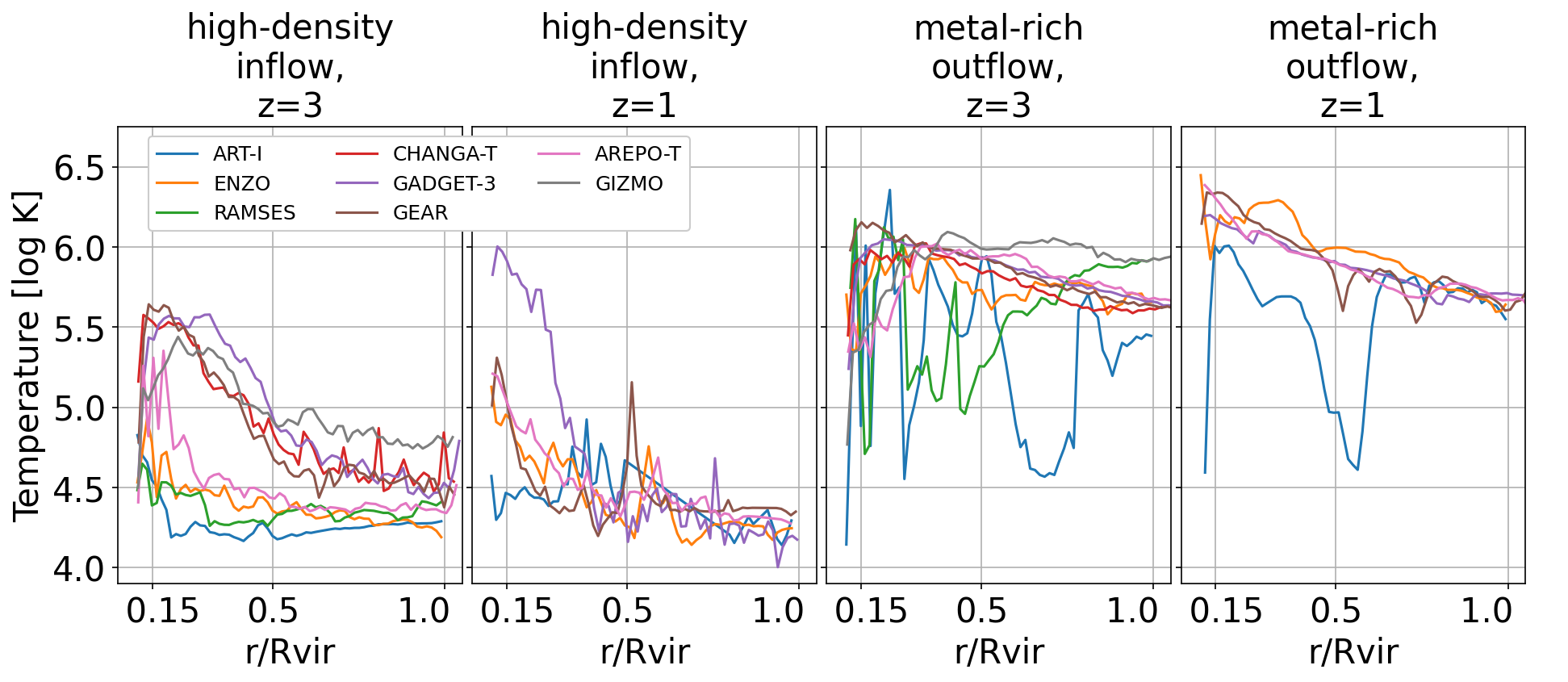}
\caption{Profiles of temperature with distance to the galaxy center as a fraction of \Rvirtext. The left two panels show profiles of dense ($n>10^{-2.5}\,\textrm{cm}^{-3}$), inflowing gas, and the right shows metal-rich ($Z>0.1Z_\odot$), outflowing gas, at redshifts \redshift{3} and \redshift{1}.}
\label{fig:Tprofile}
\end{figure*}

Another view of the inflows and outflows from the galaxy can be seen in Figure \ref{fig:Tprofile}. Here we analyze temperature profiles averaged over spherical annuli at different distances to the galaxy center. We analyze two populations of interest, which are the high-density inflows and metal-rich outflows, defined as gas parcels (cells or particles) which have $v_r<0\,\textrm{km/s}$, $n>10^{-2.5}\, \textrm{cm}^{-3}$ and $v_r>0\,\textrm{km/s}$, $Z>0.1Z_\odot$, respectively. We can see that indeed these galaxy-fueling inflows are significantly cooler than the outflows. Interestingly, there are significant differences in the profiles by code type and feedback mechanism. First, grid codes (here including \code{arepo-t}) at \redshift{3} have their fueling inflows heat up significantly less on the final approach to the galaxy than particle codes, reaching around $10^{4.5}$K to the particle codes' $10^{5-5.5}$K. This difference between code types might be due to slightly higher densities in the cool inflows in grid codes (Figure \ref{fig:projection_8codes_z3}), giving them access to faster cooling, and interestingly is different from the result of \cite{nelson_moving_2013}, which did a similar study without explicit feedback. As time evolves to \redshift{1}, several codes do not reach this threshold density of $n>10^{-2.5}\, \textrm{cm}^{-3}$ in significant parts of their CGM, leaving gaps such as in \code{art-i} at high radial distance. At the same time, only \code{gadget-3} can still be seen reaching the high temperatures mentioned above. 

The outflows in Figure \ref{fig:Tprofile} have even more substantial differences in temperature, at about an order of magnitude from $10^5$ and $10^6$ K. At \redshift{3}, five of the eight codes follow a very similar power law, mostly codes with simple thermal feedback or weaker delayed cooling (with the exception of \code{gadget-3}). \code{art-i}, on the other hand, becomes much cooler past around 0.25 \Rvirtext, while \code{ramses} and \code{gizmo}, after an initial decline with radius like the other codes, actually increase in temperature to $10^6$K as they approach the outer halo. This remains roughly the same at \redshift{1}, except that the codes just mentioned did not reach this redshift and so it gives a (misleading) appearance of further convergence.

\begin{figure}
  \centering
\includegraphics[clip,width=\linewidth]{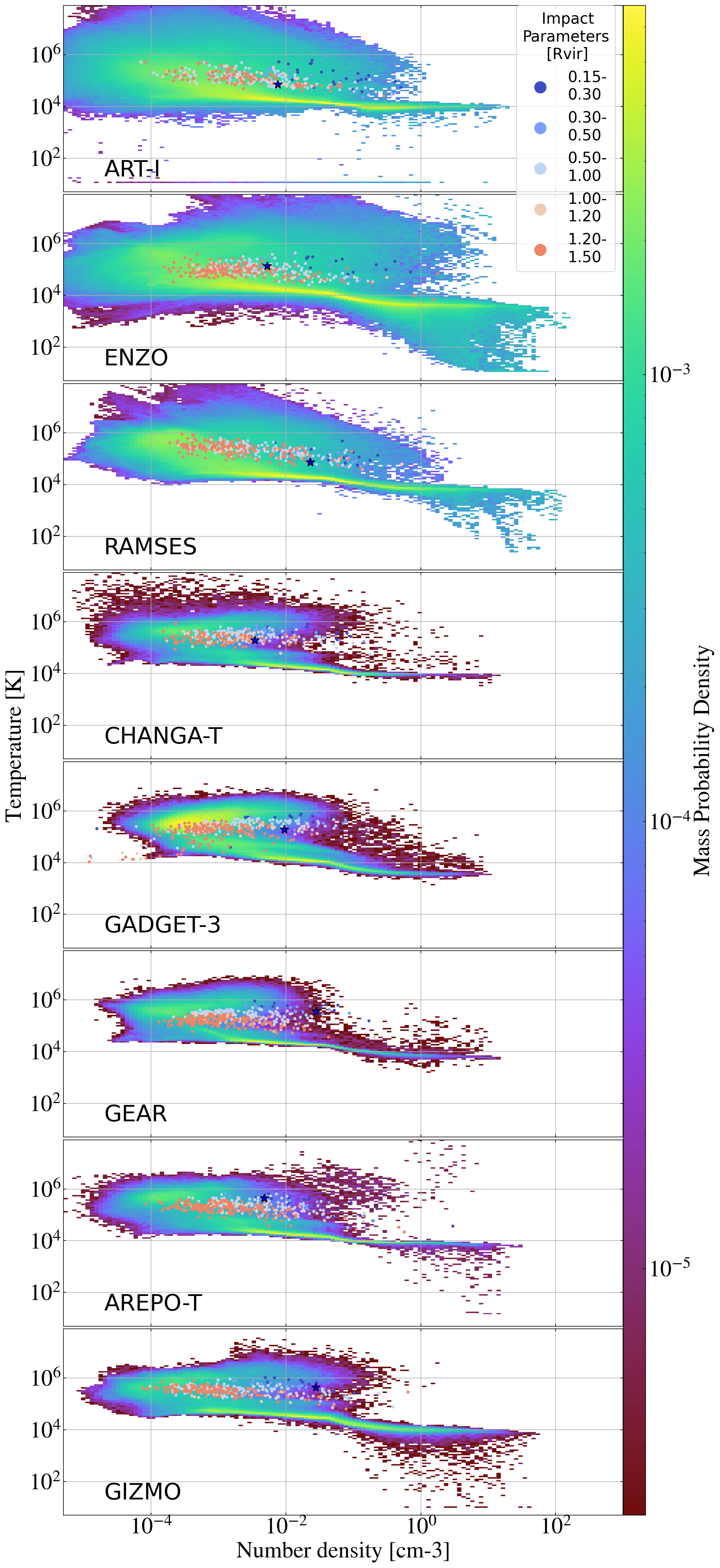}
  \caption{Phaseplot of each code at redshift \redshift{3}, showing all gas between 0.15 and 1.5 \Rvirtext (7.6 and 76 kpc). Dots indicate the average temperature and maximum density of sightlines passing through the CGM of these halos, with color indicating the impact parameter. The blue stars are the sightlines with spectra shown in Figure \ref{fig:spectra_z3}.}
  \label{fig:phaseplot_z3}
\end{figure}

\begin{figure}
  \centering
  \includegraphics[clip, width=\linewidth]{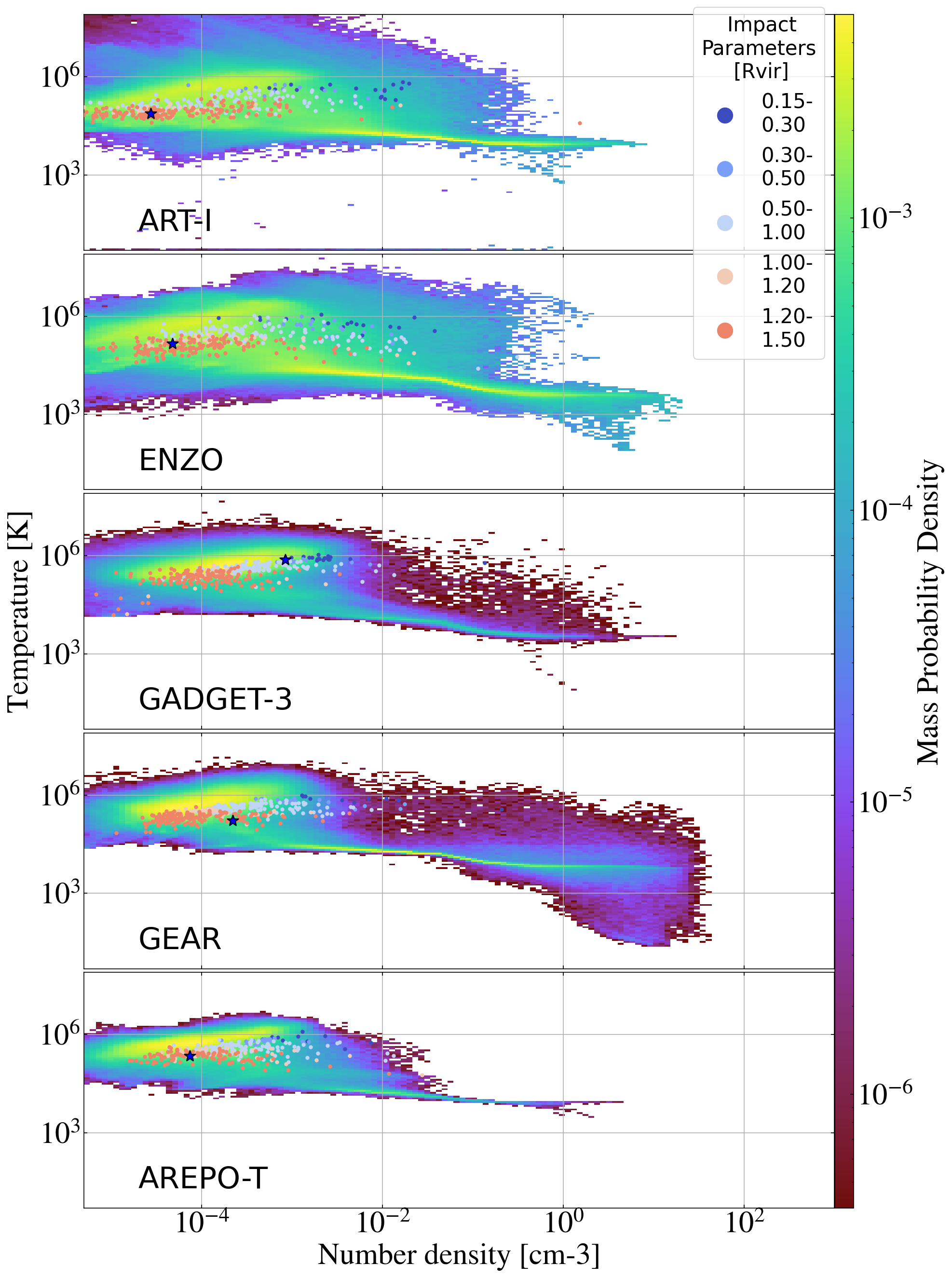}
  \caption{Identical to Figure \ref{fig:phaseplot_z3}, but at redshift \redshift{1} showing all gas between 0.15 and 1.5 \Rvirtext (23 and 230 kpc). Stars show sightlines with spectra visible in Figure \ref{fig:spectra_z1}.}
  \label{fig:phaseplot_z1}
\end{figure}

In Figure \ref{fig:phaseplot_z3}, we show the total probability density function of all gas at \redshift{3} in each of the AGORA simulations, from $r = 0.15\Rvir$ to $r = 1.5\Rvir$, thus including the CGM and some of the IGM. In all codes, a primary cooling curve is visible at around $10^4$ K. We will refer to this as the ``cooling track," which follows the minimum gas temperature for which cooling is stronger than heating (see Figure 5 in \citetalias{roca-fabrega_agora_2021}). 

There are a few interesting distinctions between the AMR and SPH type codes in Figure \ref{fig:phaseplot_z3}. AMR codes are generally more likely to fill out large clouds in phase space both above and below the cooling track, with no other really distinguishable structure. In \code{enzo}, we can even see a significant population of cold gas. SPH codes, on the other hand, have no or negligible cold gas here. They also have a much more apparent hot cloud (yellow cloud in upper left), clearly out of pressure equilibrium due to increasing in density with increasing temperature rather than decreasing. This hot cloud follows an isentropic line, meaning gas in this phase follows the equation $T n^{-2/3} = const.$, as seen in other high-temperature, low-density gas in e.g. \citetalias{kim_agora_2016} and \cite{shin_how_2021}. For gas which reaches these high temperatures, the only relevant cooling process is very slow bremsstrahlung radiation, so it then expands more or less without significant cooling. This means the particle codes have a much more straightforward two-phase structure: cool, high-density streams and hot bulk material, though to some extent this is because SPH codes do not have very many particles in the outer CGM. 

As the codes evolve to redshift \redshift{1} (Figure \ref{fig:phaseplot_z1}), they spread out to fill more of the low-density phase space, while losing most of the cold gas below and to the right of the cooling track. This may be connected to the mass threshold for virial shocks the codes cross at around this time. This mass, around $\sim 10^{12}\msun$, was first proposed in \cite{birnboim_virial_2003,dekel_galaxy_2006}, and has been explored further by a number of other groups \citep[e.g.][]{keres_how_2005,keres_galaxies_2009,faucher-giguere_baryonic_2011,van_de_voort_properties_2012,stern_maximum_2020}. In particular, in light of the results in \cite{stern_virialization_2021,hafen_hot-mode_2022}, a consequence of this shock heating of the inflowing gas could be that when the simulations reach $z<1$, gas will cool more slowly and have time to virialize (i.e. relax and rotate coherently) before entering the galaxy, with the thin gas/stellar disk forming from this coherently rotating and slowly cooling gas. While in this work we do not use redshifts below \redshift{1}, an assessment of this ``outside-in'' virialization scheme at lower redshifts will be pursued in future AGORA papers. Interestingly, the grid codes now form a similar isentropic hot cloud as mentioned for particle codes at \redshift{3} (upper left region of phase plot). This suggests that this heating effect simply takes place significantly faster in particle codes, but eventually does follow in grid codes. In all five codes, this hot phase seems to have drifted away (to lower density) from the ``cooling track''. 

On this plot, we also show the distribution of $\sim 400$ sightlines sent through the CGM, which will be examined further in Sections \ref{sec:spectra} and \ref{sec:decomposition}. The sightlines are here shown according to the density of their maximum-contribution element (where the contribution is defined as number density times path length for that element) and mass-weighted average temperature, thus showing cell features intersecting sightlines. Along a sightline, SPH codes are deposited in the form of line segments indistinguishable from grid-type ``cells"; however, this can lead to somewhat strange behavior if a sightline is far from a direct intersection with any particular gas particle, such as the extremely low-density points in \code{gadget-3}. The color indicates the impact parameter of each sightline, with blue being near the galaxy and red being at or near the virial radius. We will discuss the sightlines in more detail in the next section. The main result here is that at redshift \redshift{3}, the average temperature of sightlines remains roughly constant with increased maximum density, showing that the densest (and likely coldest) cells don't dominate the overall temperature distribution, or in other words, sightlines dominated by a high-density cell go through multiple phases with a comparable total mass contribution. There is a clear impact parameter dependence, showing more distant lines of sight are significantly less likely to go through high-density cells/regions. 

At \redshift{1}, by contrast (Figure \ref{fig:phaseplot_z1}), there is a much more significant temperature dependence on density, in both grid and particle codes. Higher-density sightlines (which remain largely close to the galaxy) have, on average, significantly hotter gas, indicating that the denser regions that lines pass through now more effectively dominate the mass distribution along the line of sight. 

\subsection{Metal Ions in Mock Spectra}
\label{sec:spectra}

\begin{figure*}
  \includegraphics[clip,width=\linewidth]{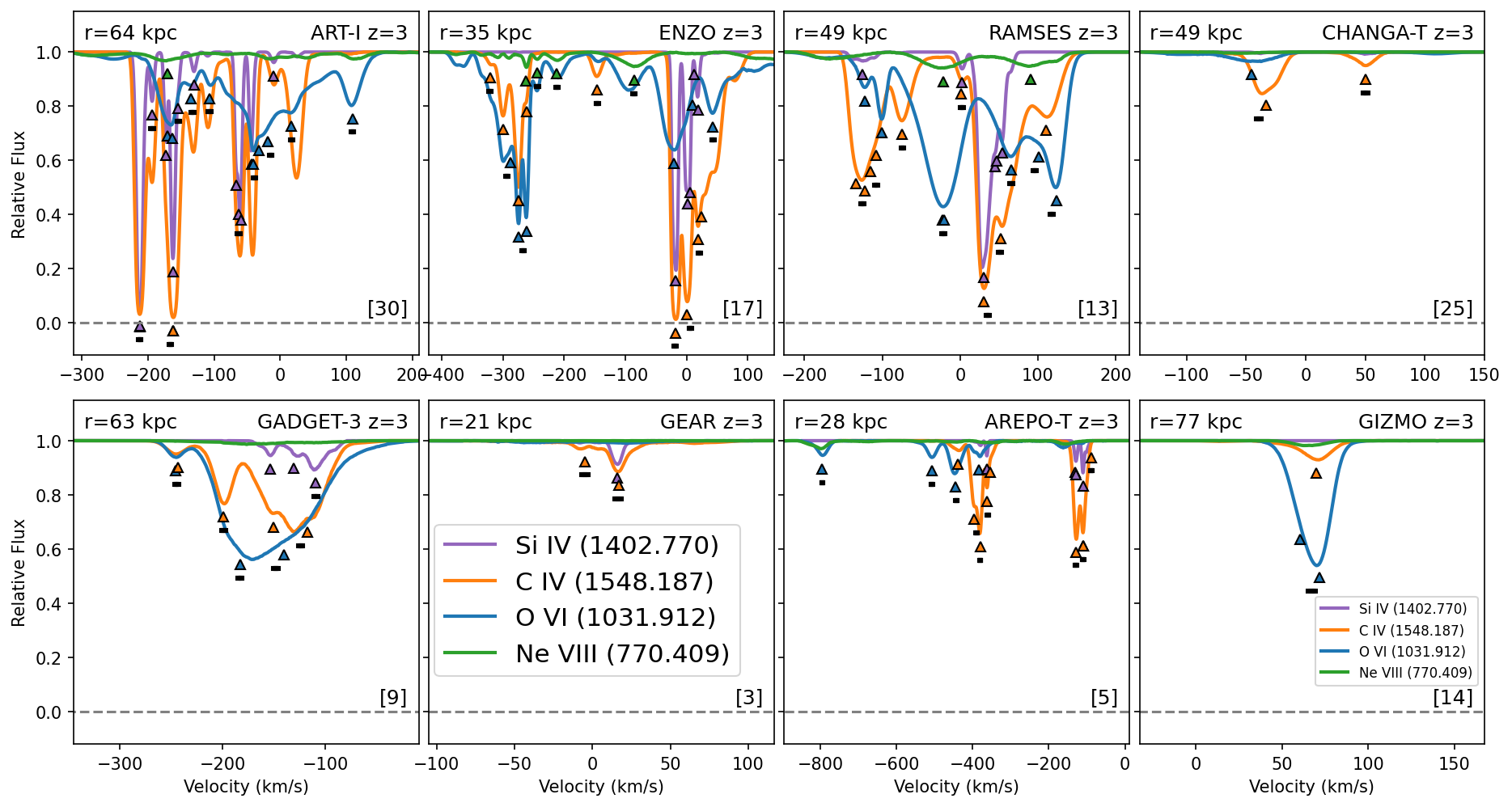}
\caption{Noiseless example spectra from snapshots of each code at \redshift{3}, here showing the strongest transition lines for medium-high ions: Si~{\sc iv}, C~{\sc iv}, O~{\sc vi}, and Ne~{\sc viii}. Triangles indicate absorption lines as detected by {\sc trident}, and black lines indicate multi-ion components, grouping together all lines found within 15 km/s of one another. Sightlines are selected by inspection to have visible components while remaining representative of 31 examined sightlines for each code. The number in square brackets indicates which line (between 0 and 30) was chosen.}
\label{fig:spectra_z3}
\end{figure*}

\begin{figure*}
  \includegraphics[clip,width=\linewidth]{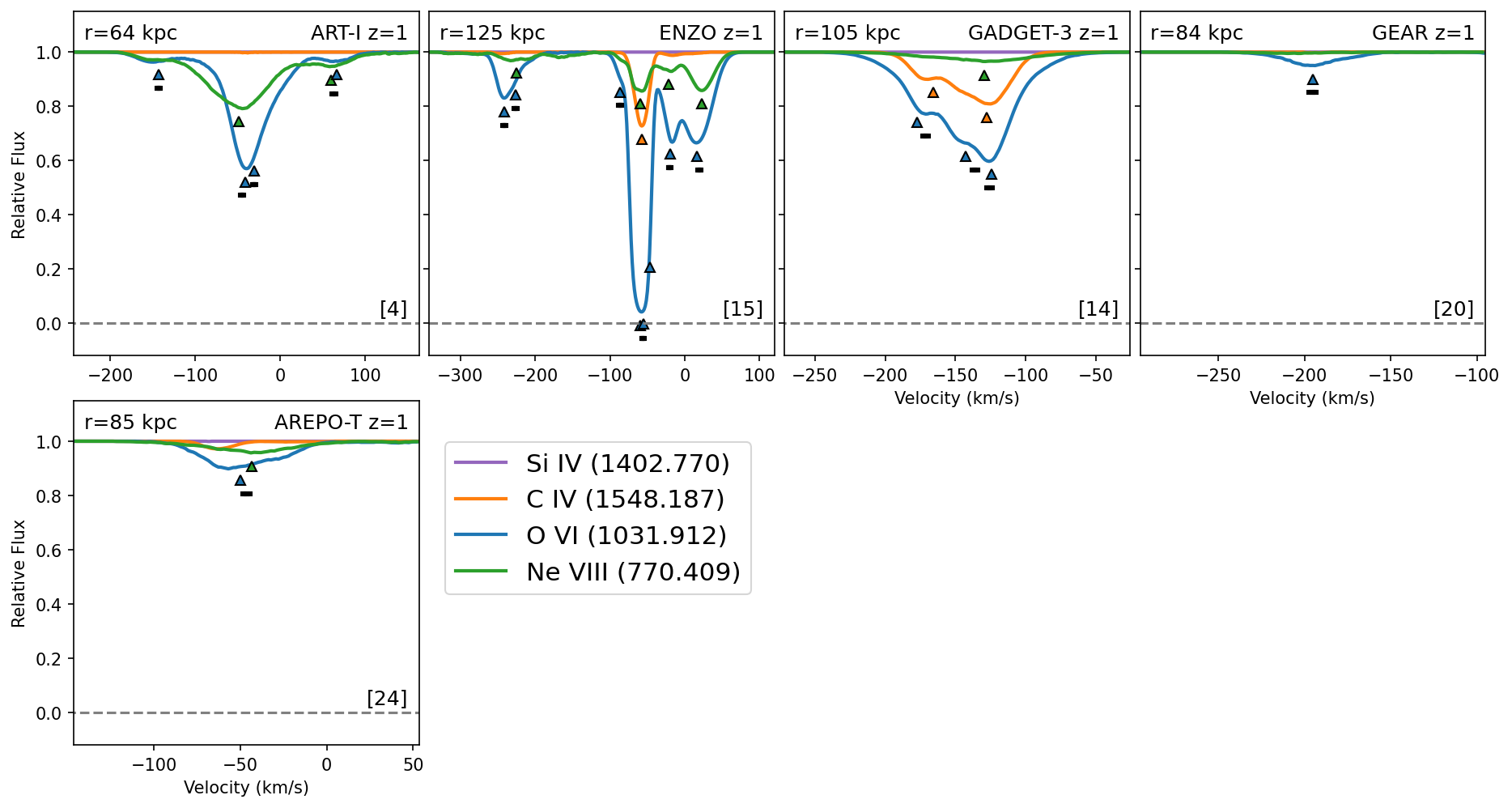}
\caption{Identical to Figure \ref{fig:spectra_z3} but at redshift \redshift{1}.}
\label{fig:spectra_z1}
\end{figure*}

\begin{figure*}
        \centering
        \includegraphics[width=0.9\linewidth]{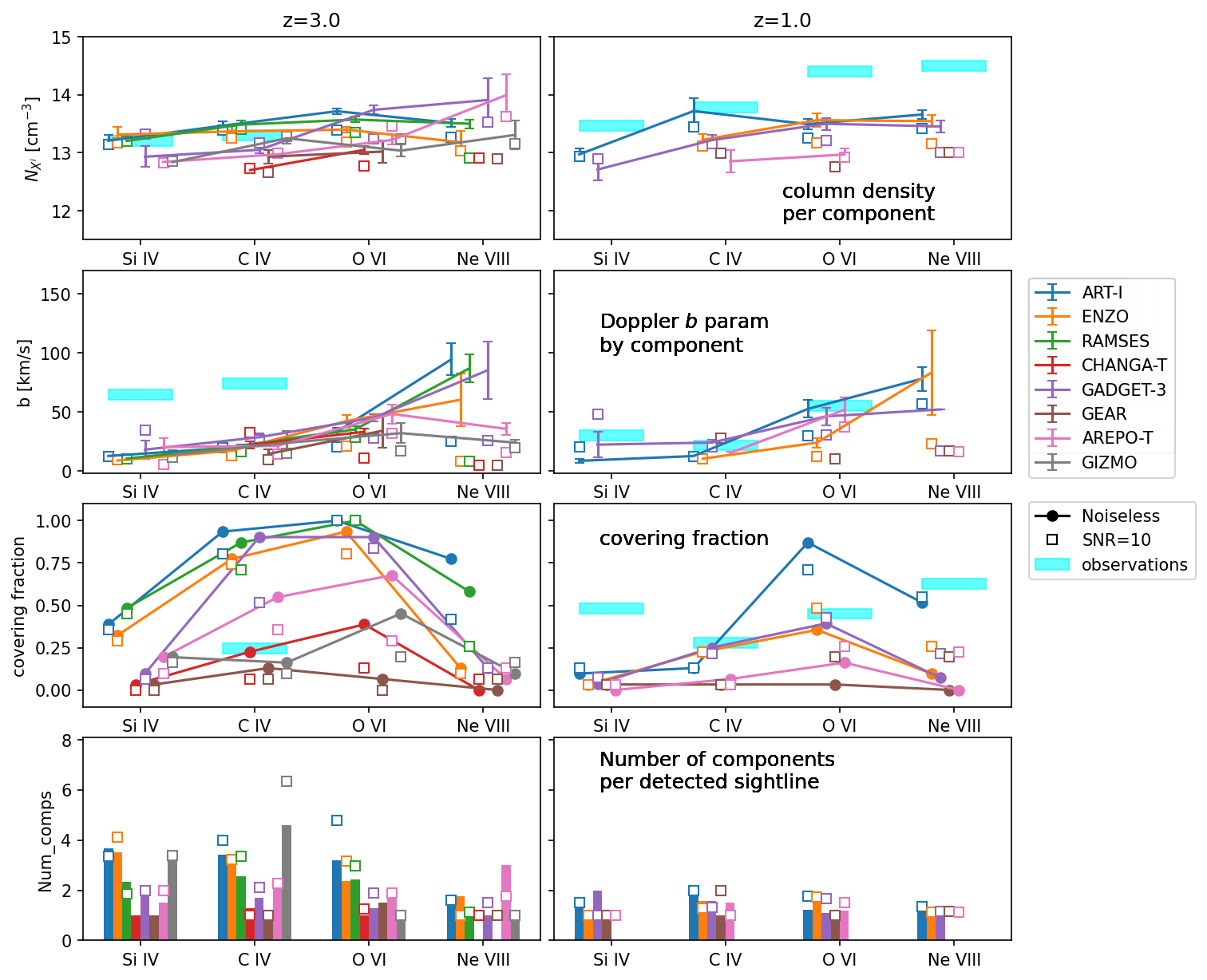}
        \caption{Analysis by ion on 31 randomized spectra through each AGORA CGM. Left column is at \redshift{3}, right column is at \redshift{1}. Colors are the same in each graph, as well as the order of small $x$-offsets added for visibility. The effects of noise on spectrum detectability are visible through comparing the noiseless results (solid markers, connected) to the results with a reasonably good S/N ratio of 10 (unfilled squares in same colors). If not enough components are detected for a particular ion in a particular code, those points are not displayed \textit{Top:} Column density per component; \textit{Second from Top:} Average Doppler $b$ parameter of each component; \textit{Second from Bottom:} Covering fraction for this ion; \textit{Bottom:} Average number of components in a sightline containing at least one component. Bright horizontal bars are estimated from observational work with arbitrary thickness for visibility (which does not represent an error bar). Specifically we show our own very rough estimates for column density per component and covering fractions, extracting data from \cite{galbiati_muse_2023} for \redshift{3} Si~{\sc iv} and C~{\sc iv}, \cite{chen_origin_2001} for \redshift{1} C~{\sc iv}, \cite{werk_cos-halos_2013} for \redshift{1} Si~{\sc iv}, \cite{tchernyshyov_cgm2_2022} for \redshift{1} O~{\sc vi} (see Figure \ref{fig:compare_to_obs_z1} caption), and \cite{burchett_cos_2019} for \redshift{1} Ne~{\sc viii}. $b$ parameters are generally not available in these papers, so those are sourced from \cite{galbiati_muse_2023} for \redshift{3} C~{\sc iv}, \cite{werk_cos-halos_2013} for \redshift{1} Si~{\sc iv} and C~{\sc iv}, and \cite{werk_cos-halos_2016} for \redshift{1} O~{\sc vi}.}
        \label{fig:spectrum_analytics}
\end{figure*}

Our understanding of the CGM in the real Universe, rather than in simulations, is generally predicated on observing different ionization levels for astronomical metals, which probe different temperature and density regions. We expect that as the ionization state depends sensitively on multiple variables (temperature, density, metallicity), the different AGORA CGMs should be very different compared to observations. In this AGORA project we categorize how each of these variables contributes to observable results, rather than attempting to track which code or feedback mechanism is ``best," though future projects could do further analysis of how well different feedback strategies fit the observations. In this section, we analyze some characteristic spectra (Figures \ref{fig:spectra_z3}, \ref{fig:spectra_z1}, and \ref{fig:spectrum_analytics}), as well as decompose ion column densities into their constituent factors (Figure \ref{fig:ion_decomposition}). We focus on four medium-high ions: Si~{\sc iv}, C~{\sc iv}, O~{\sc vi}, and Ne~{\sc viii}. These were chosen because these are the most commonly observed higher ions, generally because they have very strong lines. We avoided analysis of low ions O~{\sc ii} or Mg~{\sc ii} because none of these codes should be able to resolve the small clouds expected to host them \citep{hummels_impact_2019}. We then compare the radial column density profiles to a selection of observational results and present some insights as to what causes convergence or divergence from these results.

It is apparent that there are dramatic differences in the visible mock spectra\footnote{\code{trident}'s default behavior was modified to create spectra with LOS velocity rather than using cosmological redshifting along the line, see \url{https://github.com/trident-project/trident/pull/196}} for the selected ions in each code. In Figures \ref{fig:spectra_z3} and \ref{fig:spectra_z1}, we examine several lines at both \redshift{3} and \redshift{1}, chosen out of a sample of 31 lines to be representative of the simulation overall while containing at least some detectable absorption. Because stochastic effects would make direct comparisons of ``the same'' sightlines unlikely to probe exactly the same phases in different simulation instantiations, we instead randomize sightlines in each code independently and take this set to be a full and independent sample. Noiseless spectra are used here to more deeply understand the physical conditions underlying detections. In Figure \ref{fig:spectrum_analytics} and associated discussion we will examine the effect of adding noise to these spectra at a given signal-to-noise ratio. Voigt profiles are identified using the built-in \code{trident} line fitting tool \citep{egan_bringing_2014}, with centroids marked with triangles. Absorption lines within 15 km/s of one another are considered part of the same ``component," and components are marked with black bars. We will analyze the spectra on a code-by-code basis, also comparing the two redshifts if they are available.

\begin{itemize}
    \item \textit{ART-I}: \code{art-i} has spectra which at \redshift{3} contain both deep and wide absorption lines, with many components. While there is some amount of overlap between O~{\sc vi} and C~{\sc iv}, the lines are generally not well connected, with C~{\sc iv} much more closely tracing Si~{\sc iv}. As \code{art-i} evolves to \redshift{1}, there is an evolution towards higher ions. While absorption gets significantly weaker and broader in general, we also see that O~{\sc vi} has become the dominant line and is generally accompanied by Ne~{\sc viii}, while C~{\sc iv} has reached a negligible level. 
    \item \textit{ENZO}: Like \code{art-i}, \code{enzo} shows a large number of fairly deep and wide absorption lines in C~{\sc iv} and O~{\sc vi} at \redshift{3}, with each dominating in different components, in addition to small amounts of Ne~{\sc viii}. The main components are also quite widely separated in velocity-space, so the scale is significantly wider than all other codes besides \code{ramses}. \code{enzo} evolves to \redshift{1} by becoming weaker in general, except for growth in Ne~{\sc viii}, which is mostly aligned with O~{\sc vi}, though some C~{\sc iv}/O~{\sc vi} alignment is still visible.
    \item \textit{RAMSES}: \code{ramses} at \redshift{3} contains very wide O~{\sc vi} lines with only minimal overlap with also significant C~{\sc iv} lines. In some cases, cooler clouds are "bracketed" by presumably hotter clouds, like the two Ne~{\sc viii} components detected on either side of the deep C~{\sc iv}/Si~{\sc iv} component at $\sim 25 km/s$. This occurs regularly throughout \code{ramses} spectra.
    \item \textit{CHANGA-T}: The SPH codes generally have less absorption overall in these ions. \code{changa-t} has some clouds of both C~{\sc iv} and O~{\sc vi}, with the former generally being stronger. The two are often loosely aligned, but not perfectly, indicating they follow similar dynamics, but are generally not in the same clouds. Some clouds further show detectable Si~{\sc iv} aligned with the C~{\sc iv}, though that is not visible in this figure.
    \item \textit{GADGET-3}: In \code{gadget-3} at \redshift{3}, there is more significant absorption than in the other particle codes. Larger O~{\sc vi} components tend to be aligned with, or almost ``contain'', slightly weaker C~{\sc iv}  lines, the most significant of which also tend to contain detectable Si~{\sc iv}. This structure is only minimally changed as \code{gadget-3} approaches \redshift{1}, with the main difference being that the strongest components, rather than containing any Si~{\sc iv}, now contain a small amount of Ne~{\sc viii}, with extremely wide lines.
    \item \textit{GEAR}: \code{gear} almost never has detectable absorption in any ions except when the sightline passes through the very innermost part of the halo or the galaxy. Nevertheless, some relatively significant and deep clouds can be seen in both Si~{\sc iv} and C~{\sc iv}. O~{\sc vi} is very rare. Evolution to \redshift{1} affects mostly what species are visible. The kinds of components which previously appeared in C~{\sc iv} are now visible instead in O~{\sc vi}.
    \item \textit{AREPO-T}: \code{arepo-t} has somewhat deeper absorption lines at \redshift{3} than particle-based codes, and interestingly has an extremely wide spread in velocity-space, with multiple very discrete clouds separated by hundreds of km/s. These mostly align C~{\sc iv} with Si~{\sc iv}, or O~{\sc vi} dominated clouds even further out in velocity-space. As \code{arepo-t} evolves to \redshift{1}, it becomes significantly weaker, with Ne~{\sc viii} becoming stronger than C~{\sc iv}, and the lower ions fading.
    \item \textit{GIZMO}: The \code{gizmo} run has generally few components per sightline, though there can be significant absorption along them. In the spectrum shown in Figure \ref{fig:spectra_z3}, we again see a ``bracketing'' behavior, where two O~{\sc vi} components (which align closely enough that they give the impression of one slightly skewed line) are seen on either side of a C~{\sc iv} component. 
\end{itemize}

Next, we examine more quantitatively the properties of the absorption lines detected by \code{trident} using the methodology described in \cite{egan_bringing_2014} in Figure \ref{fig:spectrum_analytics}. Here we create spectra for 31 sightlines through each simulation, and analyze each twice. Once, using the noiseless spectra of Figures \ref{fig:spectra_z3} and \ref{fig:spectra_z1}, and then again with Gaussian noise added so that the signal-to-noise ratio (SNR) is 10, on the higher end of modern observational capacity. The noiseless results are in solid colors, and the SNR = 10 results in empty squares. Rough observational results, when available, are shown as cyan rectangles.

First, we see that, the column densities of the individual components are similar among all the codes, increasing with higher ionization energy from about $10^{12.5}$ to about $10^{13.5}$ cm$^{-2}$, with very little evolution over redshift. At \redshift{3}, this roughly agrees with observations, but at \redshift{1}, observed components are substantially larger, either due to higher noise making smaller components undetectable, or through physical divergences between the codes and observations. Similarly, the line width, or $b$ parameter, remains fairly similar between codes (though substantially below observations) at \redshift{3}. $b$ increases with increasing ionization energy for most codes, besides \code{gizmo} and \code{arepo-t}, and noise can be seen to cause decreases in width in higher ion species. At lower redshift, this conclusion remains broadly the same, though all widths are somewhat decreased compared to \redshift{3}, with observational values also falling to reach rough parity with the simulations. Since the CGM is getting hotter, as we saw above in comparing Figures \ref{fig:phaseplot_z3} and \ref{fig:phaseplot_z1}, this indicates that turbulence, the other source of Doppler broadening, must be decreasing.

The covering fractions have significantly more variation. At redshift \redshift{3}, we see that all three grid codes, and \code{gadget-3}, have more or less uniform coverage of C~{\sc iv} and O~{\sc vi}, even though those are usually used to probe very different clouds of gas. Two of them, \code{art-i} and \code{ramses} even extend this to Ne~{\sc viii}, though with somewhat less coverage. Particle codes, on the other hand, have a clear peak around O~{\sc vi}, with the exception of \code{gear} which peaks at lower ionization level with C~{\sc iv}. Noise usually decreases covering fractions, except when they are very close to 0. Interestingly, it is the lower C~{\sc iv} covering fraction in the noisy spectra of particle codes that most closely aligns with the observations \citep{galbiati_muse_2023}. Covering fractions for most ions lower as the codes evolve to \redshift{1}, however all codes moderately increase their Ne~{\sc viii} covering fraction, at least in the SNR=10 data. This shows that the CGM is generally getting hotter over time (see also Figure \ref{fig:Tprofile}). The Ne~{\sc viii} covering fraction remains noticeably higher in \code{art-i} at both redshifts, making it the only one approaching the value in \cite{burchett_cos_2019}. \code{art-i} also sees a general collapse in C~{\sc iv} and Si~{\sc iv} detections at this redshift. The most significant disagreement with observations here is in Si~{\sc iv}, which in \cite{werk_cos-halos_2013} was significantly more likely to be detected than in any code. This could be an artifact of the lower redshifts and smaller impact parameters used in COS-Halos (Figure \ref{fig:compare_to_obs_z1}), or it could result from the codes' resolution limitations having difficulty generating clouds for ions lower than C~{\sc iv}.

Completing this analysis, in the fourth row of Figure \ref{fig:spectrum_analytics} we track the total number of detected components in sightlines which were covered. In other words, if an ion is detected at least once in a sightline, how many components (usually interpreted as ``clouds'', though see \cite{marra_examining_2022} for a counterargument) is it found in.\footnote{As visible in Figure \ref{fig:spectra_z3} (e.g. \code{ramses} near -100 km/s in C~{\sc iv} and near +50 km/s in O~{\sc vi}), sometimes multiple Voigt profiles are fitted very near one another, and are thus considered part of the same ``component''. These are \textit{not} considered multiple components in the bottom row of Figure \ref{fig:spectrum_analytics}.} Generally there are more clouds detected with noise, as some noise patterns can make what is really a single component look like two peaks. There is a significant gap between grid and particle codes in the number of O~{\sc vi} components at \redshift{3}, and a smaller one in C~{\sc iv}. \code{gizmo} is an exception here, and generally shows more fragmentary components than the other particle codes. Ne~{\sc viii} almost always has a small number (1-2) of components. At lower redshift, interestingly, while the coverage increases or is maintained for O~{\sc vi} and Ne~{\sc viii}, the number of components goes down for all species in most codes, suggesting that clouds are getting bigger and more uniform, even while becoming less numerous.

\subsection{Metal Ion Origins}
\label{sec:decomposition}

While spectra can lead to useful information would be difficult to estimate with more simplistic analysis methods \citep[see for example][]{hafen_halo21_2023}, it is also useful to disentangle the source of the differences between codes more precisely. The column density of an ion can be decomposed into the product of three factors times a constant abundance $A_x$, as described in Equation \ref{eq:factors}. Often, absorption line systems are assumed to probe only one of these variables, sometimes leading to confusion or misleading statements.

\begin{figure*}
    \centering
    \vspace{0mm}
    \includegraphics[width=0.923\linewidth]{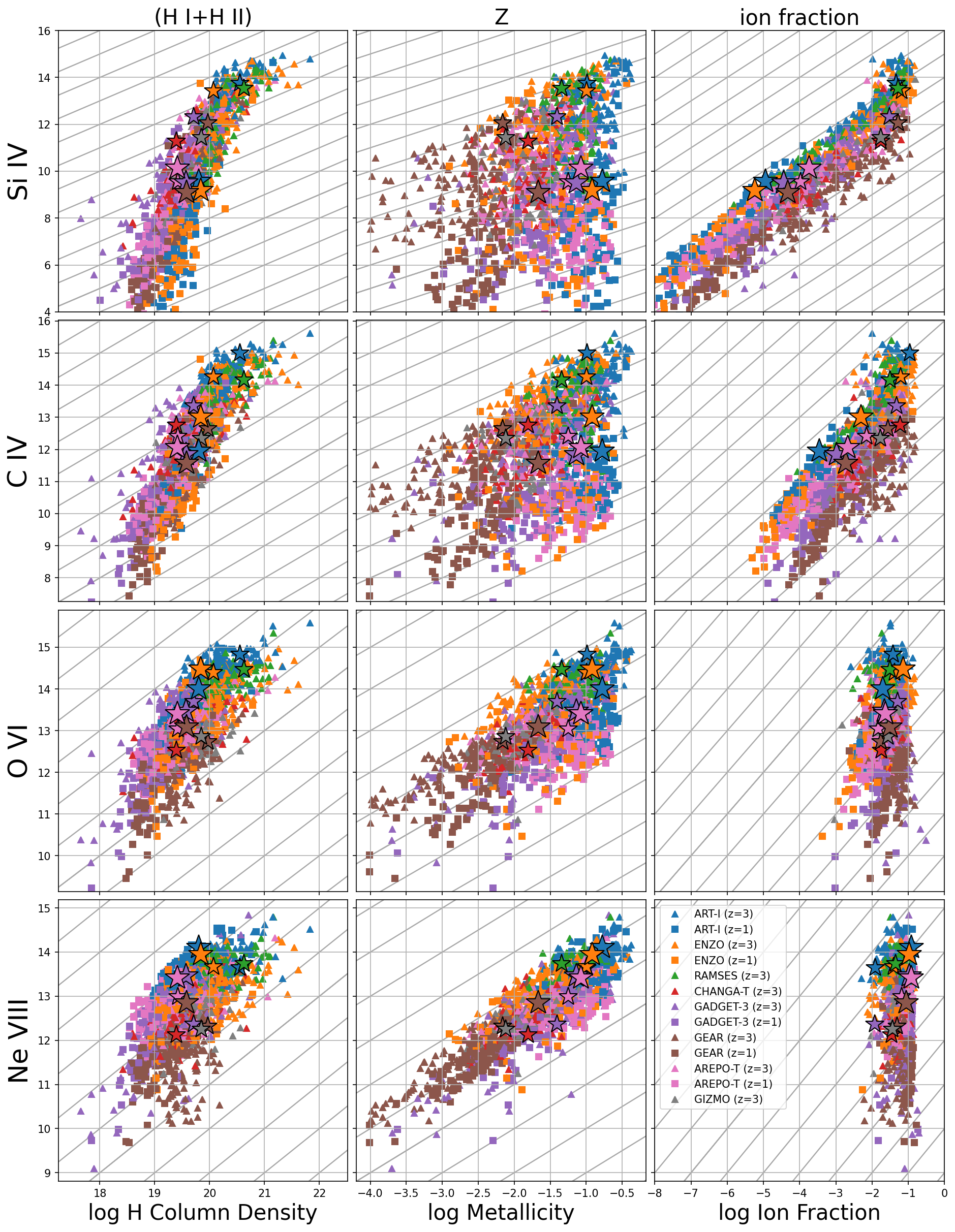}
    \vspace{0mm}
    \caption{Origin of ions along sightlines. Columns adjust whether the $x$-coordinate is hydrogen column density, sightline metallicity, or ion fraction. Rows have $y$-coordinates as column density of Si~{\sc iv}, C~{\sc iv}, O~{\sc vi}, or Ne~{\sc viii}. These ions are sorted by increasing ionization energy from top to bottom. Colors indicate different codes from the AGORA simulation, and different shapes indicate different redshifts (\redshift{3}, triangles, and \redshift{1}, squares). Smaller and larger stars show the selected line in each code at \redshift{3} and \redshift{1} highlighted in Figures \ref{fig:spectra_z3} and \ref{fig:spectra_z1}, respectively.}
    \label{fig:ion_decomposition}
\end{figure*}

In Figure \ref{fig:ion_decomposition}, we examine this situation by separating out the three variables. Here we show a suite of $\sim$ 400 lines of sight passing through each galaxy's CGM. For each of the same four ions, Si~{\sc iv}, C~{\sc iv}, O~{\sc vi}, and Ne~{\sc viii}, we have directly calculated the column density along each line of sight. For grid codes, this is the sum of ion number density (calculated with \code{trident}) times sightline path length for each cell in the sightline path. For particle codes, column density is instead calculated by dividing the path length into discrete sections defined by the smoothed gas particle field, and then integrating the ion number density of that smoothed particle times section length.\footnote{see Turk et al. in prep for details on how \code{yt} and therefore \code{trident} have been updated to handle particle codes.} These column densities are the $y$-values of the points in the scatterplots of figure \ref{fig:ion_decomposition}, with the same sightlines appearing in each panel.

The column densities described above are plotted against the total hydrogen column density (left, calculated similarly to the ion number densities), average metallicity (center, calculated as the total metal column density over total hydrogen column density), and the total ion fraction along the LOS (right, calculated as the column density of the given ion divided by the total column density for the element). The diagonal lines on each image are linear relationships, so if one factor alone could explain the variation in column density, points would follow these lines in one column, and have no correlation in any other column. To guide the eye, and for comparison to available spectra, lines which were highlighted in Figures \ref{fig:spectra_z3} and \ref{fig:spectra_z1} are plotted as small and large stars, respectively.

What is remarkable about this factorization of column density is that there is no single variable that controls ion column densities. Instead the relationship appears to change with ionization energy, with lower ions (upper rows) following a different pattern than higher ions (lower rows). 

The lower ions Si~{\sc iv} and C~{\sc iv} appear to track much more strongly with ion fractions than anything else, and the higher ions O~{\sc vi} and Ne~{\sc viii} instead track most closely with metallicity. There appears to be a continuous morphing of the shapes in both the center and right columns as one tracks from the top row to the bottom. The center column compresses from a wide scatter to a linear relationship along the $N_{X^i} \propto Z$ lines, while the rightmost column starts as a clear linear relationship for low ions and flattens out into an approximately constant $f_{X^i} \simeq 0.1$ for high ions. The leftmost column is less clear, because ion fraction depends sensitively on density so these values are not independent. While this image only shows four ions, this trend remains uniform to both higher (e.g. Mg~{\sc x}) and lower (e.g. Mg~{\sc ii}) ionization states besides the ones shown here. 

The multiple simulations with controlled conditions used in the AGORA project are useful for our interpretation of this result. For example, let us compare the distribution of sightlines by code (color). In the center column of Figure \ref{fig:ion_decomposition}, metallicities are tightly grouped together on a code-by-code basis by color, especially in the bottom row. In the ion fraction graphs, however, all codes follow very similar tracks, with wide spread in ion fraction for low ions, and a very narrow range for high ions. Thus, ion fraction depends more strongly on the sightline position within the simulation (for low ions), while metallicity depends more on the parameters of the simulation itself.

If rather than a large number of calibrated simulations, we were only studying one or two implementations, it would have been very straightforward to see intra-code ion fraction variations, and much harder to see inter-code metallicity variations. If two implementations were close in metallicity, this variable would appear to have negligible impact, and if they were widely separated, it would appear to simply make the codes impossible to directly compare. Only with a large number of codes that fill in a wide phase space of possible metal diffusion patterns, as in AGORA, is the increasingly linear relationship with increasing ionization potential between metallicity and column density visible. Most uncalibrated simulation suites would struggle to disentangle confounding effects such as differences in mass and environment, whereas here the physical reason would be more likely to relate to the implementation of the ionization models within \code{Cloudy}under the variety of conditions caused by different systems of metal diffusion and models of feedback.

\begin{figure}
  \includegraphics[clip,width=\linewidth]{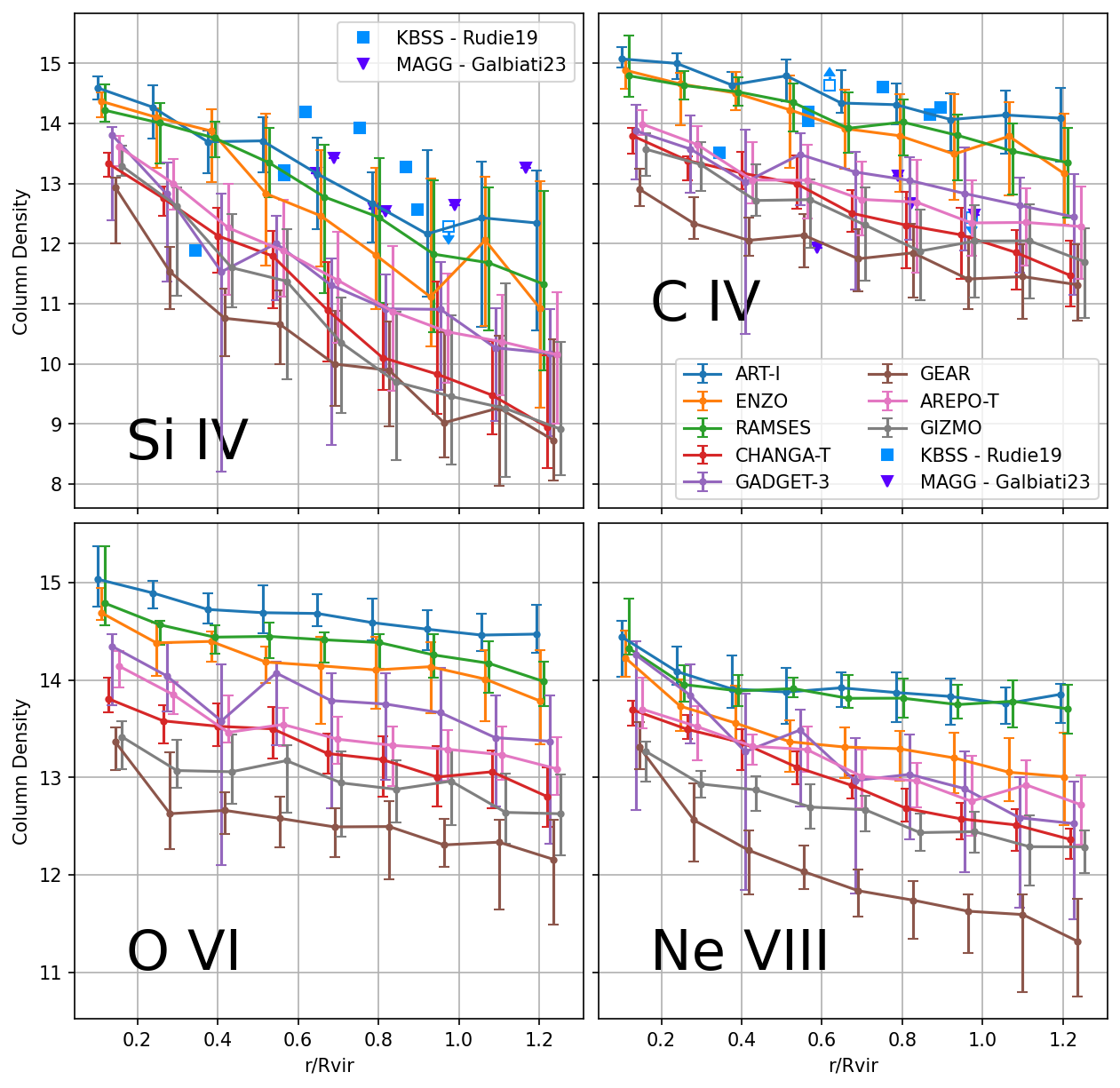}
  \caption{Comparison of radial column density profiles between AGORA galaxies and relevant observations at \redshift{3}. Non-detections and saturated lines are indicated with open squares, with a downward or upward arrow, respectively. In this figure, points labeled ``KBSS - Rudie19" and ``MAGG - Galbiati23'' are taken from \cite{rudie_column_2019} and \cite{galbiati_muse_2023}.}
  \label{fig:compare_to_obs_z3}
\end{figure}

\begin{figure}
  \includegraphics[clip,width=\linewidth]{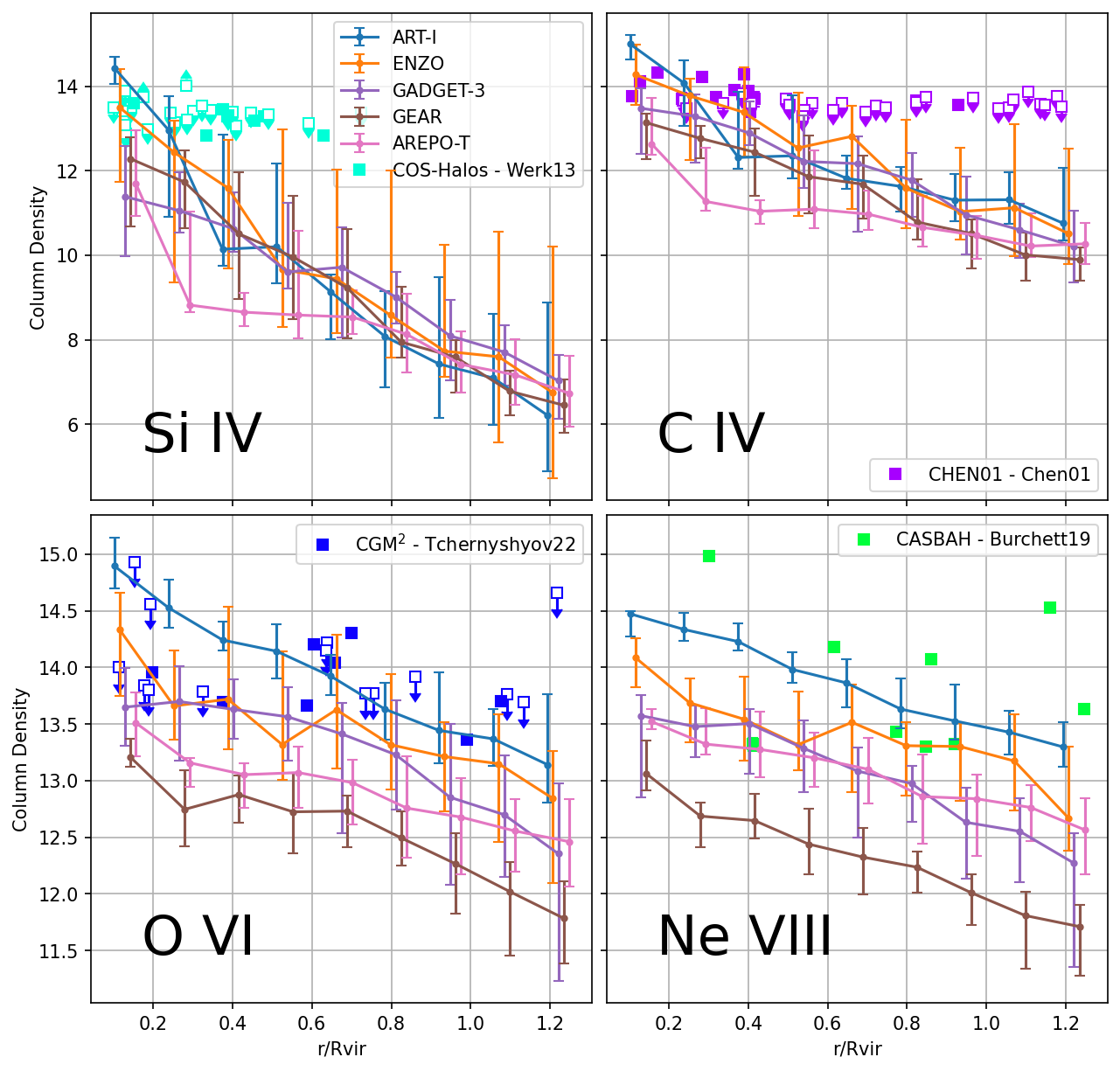}
  \caption{Identical to Figure \ref{fig:compare_to_obs_z3}, but at \redshift{1}. In this figure, points labeled ``COS-Halos - Werk13" are taken from \cite{werk_cos-halos_2013}, ``Chen01" are from \cite{chen_origin_2001}, ``CGM$^2$ - Tchernyschov22" are from \cite{tchernyshyov_cgm2_2022}, and ``CASBaH - Burchett19" are from \cite{burchett_cos_2019}. The latter two surveys are closer to \redshift{1}, with \cite{tchernyshyov_cgm2_2022} having sufficient data to filter by redshift, so here we show only points with $0.4 < z < 1.0$. The former two were at lower redshift ($z<0.4$) and so are only approximately comparable to the AGORA galaxies.}
  \label{fig:compare_to_obs_z1}
\end{figure}

\subsection{Comparison with Observations}\label{sec:compare_to_obs}

In Figures \ref{fig:compare_to_obs_z3} and \ref{fig:compare_to_obs_z1} we see that there are significant differences in the radial profiles of ion column densities in the different simulations compared to observations. The connected dots represent the median column density values at that distance, and the error bars are the 16th and 84th percentiles over the same 400 sightlines used in Section \ref{sec:spectra}, which would correspond to one standard deviation if the column densities followed a Gaussian distribution (which they generally do not). We will point out that this comparison is inherently limited by the difference between the simulated and observational datasets. The observations chosen were designed to be in the CGM of similarly-sized galaxies at around the same redshift, but will inherently measure unknown phases through the CGM of distinct halos, while the simulation data is, for each snapshot, multiple sightlines through the same halo. This inherently means that while some variation naturally does reflect the random phases the sightlines may pass through, other variation will reflect the different halos and environments, which is not available in AGORA. The relatively small error bars on the simulation data show that even though the CGM is a multiphase medium, the different phases are distributed in such a way that most sightlines sample many available phases, and so different lines of sight with the same impact parameter have similar column densities. However, it is clear that these distributions can be very different between codes. Because the CGM is relatively unconverged between codes according to multiple metrics (gas temperature, density, metal distribution, and to some extent, resolution), it is not recommended to interpret these results as primarily indicating which feedback system (or which codes) agree ``most closely" with observations. Rather, what is most useful about this analysis is to disentangle which metrics matter more for the ion of interest. 

For example, at \redshift{3} (Figure \ref{fig:compare_to_obs_z3}) we can see that there is a very clear bimodality between the grid and particle type codes, which is most visible for the Si~{\sc iv} and C~{\sc iv} profiles. For Si~{\sc iv} and C~{\sc iv}, the grid codes are more or less aligned with the data in \cite{rudie_column_2019, galbiati_muse_2023} where there is data outside the innermost halo,\footnote{\label{footnote:EW_to_N}The data in \cite{galbiati_muse_2023} is generally reported as equivalent widths rather than column densities. We convert to column density here and in Figure \ref{fig:spectrum_analytics} using Equation 2 of \cite{ellison_sizes_2004},
\begin{equation}
    N=1.13\cdot10^{20}\frac{EW}{\lambda_0^2 f},
\end{equation}
where $N$ is the column density in cm$^{-2}$, $EW$ is the component equivalent width in \AA, $\lambda_0$ is the rest wavelength of the transition in \AA, and $f$ is the oscillator strength of the transition, taken from \code{trident} documentation. This equation requires the profile to be in the linear regime, meaning $EW<0.2$ \AA. We get relative distances by dividing \cite{galbiati_muse_2023} impact parameters in kpc by the AGORA \redshift{3} virial radius, 53 kpc.} with still some slight underprediction for Si~{\sc iv} at mid-range (0.5 -- 0.8\Rvirtext). It is notable that the higher ions remain more constant with impact parameter, especially at higher distances from the CGM. This makes sense considering that higher ions are more sensitive to metallicity than gas state as shown in Figure \ref{fig:ion_decomposition}, which depends more on which code is used than where the sightline penetrates it due to differences in metal mixing and diffusion.

As the codes evolve to \redshift{1} (Figure \ref{fig:compare_to_obs_z1}), there is a significant convergence in the Si~{\sc iv} and C~{\sc iv} profiles, while the O~{\sc vi} and Ne~{\sc viii} profiles remain more spread out over three orders of magnitude. All codes drop much lower than the detectability threshold within 0.3 \Rvirtext for Si~{\sc iv} \citep{werk_cos-halos_2013} and C~{\sc iv} \citep{chen_origin_2001},\footnote{\cite{chen_origin_2001} also reports equivalent width instead of column density, see footnote \ref{footnote:EW_to_N}.} while only \code{art-i} and \code{enzo} seem to generate enough metals to match the O~{\sc vi} profile from \cite{tchernyshyov_cgm2_2022} in the outer halo (though more codes are close in the inner part of the halo). Ne~{\sc viii} has a much more significant scatter in \cite{burchett_cos_2019}, and no code really effectively resolves it; however, the scatter in the simulations remains fairly low, indicating perhaps that metal mixing is too efficient (as we can see with the high degree of homogeneity in sightlines by code in Figure \ref{fig:ion_decomposition}) or that the Ne~{\sc viii} dominant phase is too efficiently distributed throughout the CGM. It could also imply that the difference between different halos with different masses, environments, and other conditions, is more significant than the difference between sightlines, which would resonate with this study but unfortunately cannot be tested with the single implementations of each galaxy used in AGORA.

\begin{figure*}
        \centering
        \includegraphics[width=\linewidth]{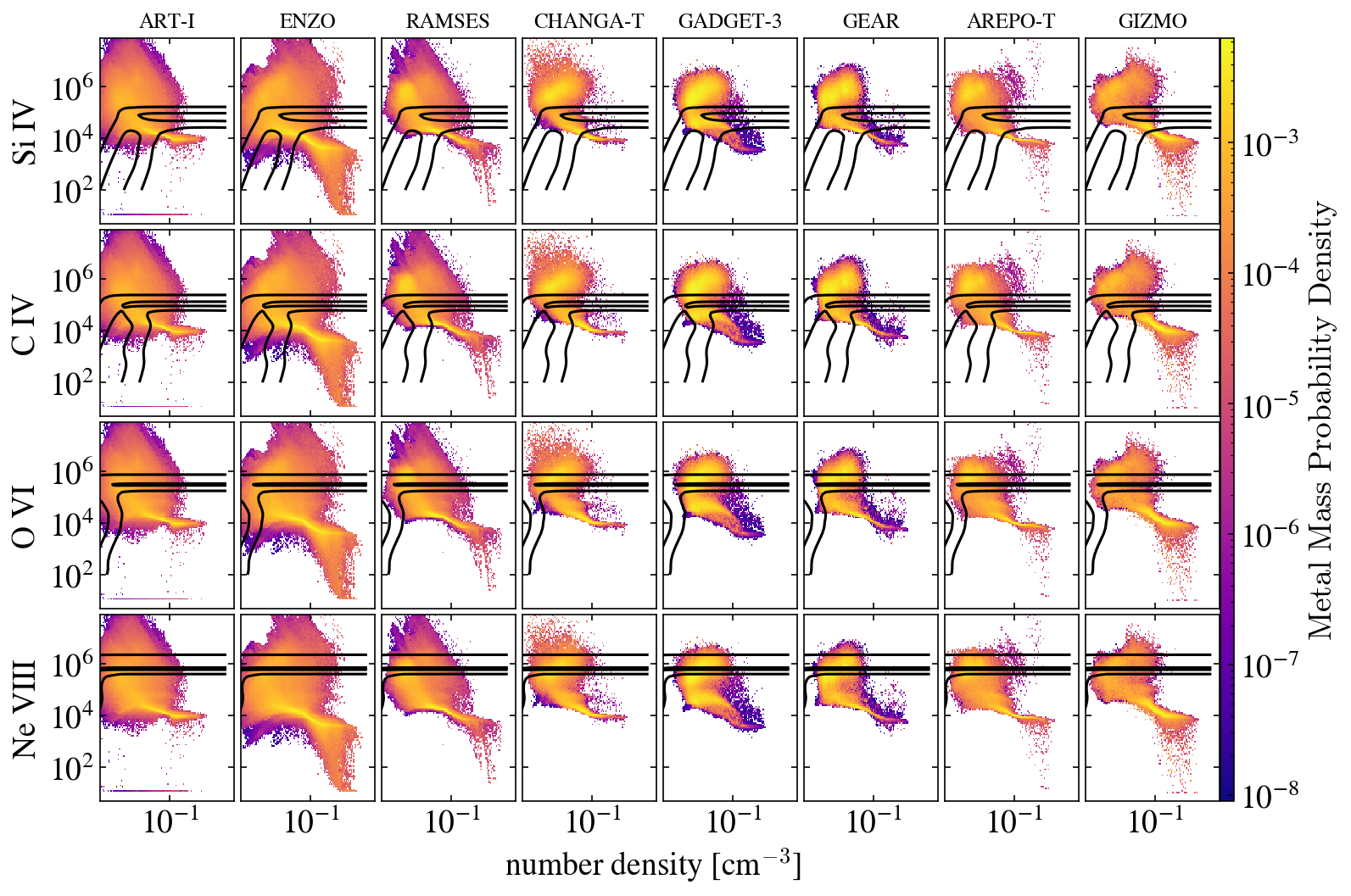}
        \caption{Map of phaseplots of all codes at \redshift{3}, similar to Figures \ref{fig:phaseplot_z3} and \ref{fig:phaseplot_z1} but colored by the metal mass, rather than the total. Columns are each code, repeated four times. Overplotted are 20 percent and 1 percent contours for each ion.}
    \label{fig:ion_mechanisms}
\end{figure*}

Finally, we showcase a relevant effect which might be causing low and high ions to respond differently to ion fraction versus metallicity, to motivate future work in this field. In Figure \ref{fig:ion_mechanisms} we show \redshift{3} phaseplots similar to those seen in Figure \ref{fig:phaseplot_z3}, except now colored by metal mass rather than total mass. Each phaseplot is repeated four times vertically, and plotted over each are the 1 percent and 20 percent ion fraction contours for the four ions being analyzed in this work. As argued in \cite{strawn_o_2021} and \cite{strawn_distinguishing_2023}, in the horizontal ``upper" ridge, each ion should be considered collisionally ionized (CI), and the diagonal ``lower" ridge, it should be considered photoionized (PI). 

As we can see here, all of these codes have their Si~{\sc iv} PI ion fraction peak either somewhat overlapping or at least near the general ``cooling track" curve, with slightly less overlap for C~{\sc iv}. O~{\sc vi} and Ne~{\sc viii} have PI peaks \citep[and in other models with different parameters, these can be important, see for example][]{stern_does_2018,strawn_o_2021}, but they take place at densities so low that they are not occupied on these phaseplots. It is important to note that the CI ion fraction peaks are not at the same temperatures for all ions. For high ions, these are approaching the bulk of the metal mass in the hot phase, while for lower ions, the collisional peak is in the less occupied ``middle'' region between the cool and hot phases. Therefore, high ions are much more ubiquitously created through collisional ionization and therefore more weakly sensitive to density. Nevertheless we note that these collisionally ionized column densities are by no means totally independent of density, as seen in the leftmost column of Figure \ref{fig:ion_decomposition}.

Examining these results, we posit that the evolution in ion factorization shown in Figure \ref{fig:ion_decomposition} from low to high ions might be correlated with the switch from dual contributions of photoionization and collisional ionization for lower ions to collisional ionization dominance for higher ones, though more research on this point will be needed and in a larger parameter space than that swept out by AGORA. These results could be substantially changed with the inclusion of more physics allowing for more small, cool clouds to survive in the halo or be created there, such as magnetic fields \citep{nelson_resolving_2020} or higher resolution \citep{peeples_figuring_2019,hummels_impact_2019,van_de_voort_cosmological_2019,ramesh_zooming_2023}. Future AGORA projects which include these improvements, as have been suggested, would be an excellent way to disentangle these effects, and possibly modify the conclusions found here.

\section{Discussion and Conclusion} 
\label{sec:conclusion}

The AGORA project is and remains primarily a community of scientists attempting to understand whether the results of cosmological and galaxy simulations are at this time converged, and what aspects of this theoretical project are and are not well understood. Scientific programming is generally not designed to be highly scalable, or to be adopted en masse and maintained by large, professional companies. Indeed, as new techniques are developed and processing power increases, scientific codes need the flexibility of being developed by a small group to remain cutting-edge enough for original research, with new codes arising whenever their need becomes apparent. Thus, a large number of groups are developing more or less redundant codes which all attempt to answer the same question: does application of known and commonly accepted galactic astrophysics create adequately realistic galaxies? It is much more rarely asked, does the application of this shared physics always create the same results with each different implementation method? AGORA was founded to analyze this question, and to generally get the backend simulation developers in contact with one another, so their simulations could be mutually intelligible.

In \citetalias{kim_agora_2013} and \citetalias{kim_agora_2016}, this question was approached by development of all codes to accept common input files which standardized the presentation of initial conditions, heating/cooling functions, visualization tools, and other aspects. With CosmoRun (\citetalias{roca-fabrega_agora_2021} and IV), it was further asked whether different (commonly used) physics prescriptions change simulation results, holding everything else, even particularities like initial conditions, constant. It was necessary to expand the scope in this way because the codes were so particularized in their development that it would be impossible to effectively modify the codes to use the same ``feedback'' (which here is only stellar feedback, though AGORA will be developing new AGN simulations in the near term) without changing the codes so dramatically from their normal use that it no longer represented a comparison between commonly used codes. This new approach made the AGORA project much more complex, as now two variables, code implementation and feedback prescription, control the outcomes instead of only one, and these outcomes are correspondingly much more different from each other than they were for the simulations in \citetalias{kim_agora_2013} and \citetalias{kim_agora_2016}. 

The result in \citetalias{roca-fabrega_agora_2021} was that even with these significant differences, the codes could be compatible with overall results in star formation, i.e. realistic star formation histories were compatible with many different feedback implementations. But as we show in Paper IV and here, other effects such as merger timing discrepancies and especially the quantity and state of mass and metals distributed into the CGM, and the state of that gas with respect to observable quantities, is vastly different, making direct comparisons more challenging. This more complex simulation space leads to significant benefits as well as challenges. Particularly, it allows us to examine a vast parameter space in a way that the individual implementation of each code or the multiple formation histories of different galaxies can be neglected, which could help us reach a more sophisticated understanding of the physics, either of the simulations or of their accompanying analysis tools.

The main results presented in this paper are as follows:

\begin{enumerate}
\item All codes retain similar total gas mass into the CGM from \redshift{6} and below, but send vastly different metal masses into this region.
\item All codes mix metals between inflowing and outflowing phases in similar ways, but they are mostly distinguished in how many metals are in either phase, according to the variety of feedback prescriptions used.
\item All codes have some amount of hot, metal-rich biconical outflows and cool inflowing streams. The outflows are significantly faster in grid codes and slower in particle codes, with moving mesh codes somewhere in between.
\item Spectra between medium-high ions are often kinematically distinct from each other, and in some codes O~{\sc vi} aligns with C~{\sc iv}; in others O~{\sc vi} with Ne~{\sc viii}, and in others no alignments are found, showing that the ions visible in spectra do not always arise from the same gas temperature-density phase.
\item Low ions are more strongly determined by ion fraction, while high ions are more strongly determined by metallicity. This difference may have to do with the photoionized or collisionally ionized origins of the species at different energy levels.
\item Most codes underpredict ion column densities for most ions, with significant spread between codes. Low ion column densities generally have more impact parameter dependence than high ions, which have stronger code and feedback type dependence instead but change less steeply with radius.
\end{enumerate}

Future work with the CosmoRun galaxies will involve more detailed comparisons with observations using the radiative transfer code \code{skirt} \citep{baes_skirt_2015}, and possibly a final follow-up on halo evolution (Papers III and IV) down to \redshift{0}. Other projects will include continuing analysis of ionization states in the CGM and further analysis of the satellite galaxies in a follow-up to Paper V. Additionally, new codes such as \code{swift} \citep{schaller_swift_2023} and \code{gadget-4} \citep{springel_gadget-4_2022} have expressed interest in joining this project. These will be added to future CosmoRun papers, though they had not finished running at the time this work was submitted. Finally, a re-run of the CosmoRun simulation with higher resolution might be executed to compare how the increased resolution changes each code, as well as allowing us to compare more detailed structures such as clumps or smaller clouds in the CGM.

Besides these, AGORA will continue to run new simulations, including simulations of an AGN interaction with the isolated disk conditions of \citetalias{kim_agora_2016}, and technical analyses of the codes' responses to heating and cooling curves (Revaz et al., in prep.). As the simulation community continues to add newer and more efficient physics and implementations, collaborators are committed to planning new AGORA simulations to continue to dive into their effects, as simulation groups around the world try to converge on all the critical questions surrounding galaxy and cosmological evolution.

\section*{Acknowledgements} 
\label{sec:acknowledgements}

We thank all of our colleagues who participate in the AGORA Project for their collaborative spirit, which has allowed the AGORA Collaboration to remain strong as a platform to foster and launch multiple science-oriented comparison efforts. We thank the UCSC Foundation Board Opportunity Fund for supporting the AGORA project papers as well as the AGORA annual meetings. We thank Volker Springel for providing the original versions of \code{gadget-3} to be used in the AGORA Project. Analysis of all codes was done using resources of the National Energy Research Scientific Computing Center (NERSC), a user facility supported by the Office of Science of the U.S. Department of Energy under contract No. DE-AC0205CH11231. Partial support for C.S. was provided by grant HST-AR-14578 to J.R.P. from the STScI under NASA contract NAS5-26555 and from J.R.P.'s Google Faculty Research Grant. C.S. also received support from the UCSC Science Internship Program (SIP) and the ARCS Foundation. 

S.R.-F. acknowledges support from a Spanish postdoctoral fellowship, under grant No. 2017-T2/TIC-5592. His work has been supported by the Madrid Government (Comunidad de Madrid–Spain) under the Multiannual Agreement with Complutense University in the line Program to Stimulate Research for Young Doctors in the context of V PRICIT (Regional Programme of Research and Technological Innovation). He also acknowledges financial support from the Spanish Ministry of Economy and Competitiveness under grant Nos. AYA2016-75808-R, AYA2017-90589-REDT, and S2018/ NMT-429, and from CAM-UCM under grant No. PR65/1922462. The \code{art-i} simulations were performed by S.R.-F. on the BRIGIT/ EOLO cluster at Centro de Proceso de Datos, Universidad Complutense de Madrid, and on the STÓCATL supercomputer at Instituto de Astronomía de la UNAM. The \code{ramses} simulations were performed by S.R.-F. on the MIZTLI supercomputer at LANCAD, UNAM, within the research project LANCAD-UNAM-DGTIC-151 and on Laboratorio Nacional de Supercómputo del Sureste, CONACYT. J.K. acknowledges support by Samsung Science and Technology Foundation under Project Number SSTF-BA1802-04. His work was also supported by the National Research Foundation of Korea (NRF) grant funded by the Korea government (MSIT) (No. 2022M3K3A1093827 and 2023R1A2C1003244). His work was also supported by the National Institute of Supercomputing and Network/Korea Institute of Science and Technology Information with supercomputing resources including technical support, grants KSC-2020-CRE-0219, KSC-2021-CRE-0442 and KSC-2022-CRE-0355. A.G. would like to thank Ruediger Pakmor, Volker Springel, Matthew Smith and Benjamin Keller for help with \code{arepo} and \code{grackle}. The \code{arepo} runs were carried out by A.G. on the High Performance Computing resources of the FREYA cluster at the Max Planck Computing and Data Facility (MPCDF) in Garching operated by the Max Planck Society (MPG). A.L. acknowledges funding by the MIUR under the grant PRIN 2017-MB8AEZ. K.N. acknowledges support from MEXT/JSPS KAKENHI grant Nos. 19H05810, 20H00180, and 22K21349, as well as travel support from Kavli IPMU, World Premier Research Center Initiative, where part of this work was conducted. The \code{gadget3-osaka} simulations and analyses were performed by K.N. and I.S. on the XC50 systems at the Center for Computational Astrophysics of the National Astronomical Observatory of Japan, on OCTOPUS and SQUID at the Cybermedia Center of Osaka University, and on Oakforest-PACS at the University of Tokyo as part of the HPCI System Research Project (hp200041, hp210090, hp220044, hp230089). Y.R. acknowledges support by the Swiss Federal Institute of Technology in Lausanne (EPFL) through the use of the facilities of its Scientific IT and Application Support Center (SCITAS). H.V. acknowledges support from PAPIIT of Universidad Nacional Autónoma de México (UNAM) under grant No. IN101918 and also from Centro Nacional de Supercomputo (CNS-IPICYT-CONACYT) and from the Laboratorio Nacional de Superc\'omputo del Sureste (LNS-CONAHCYT). The CHANGA simulations were performed by H.V. and J.W.P. on the ATÓCATL supercomputer at Instituto de Astronomía de la UNAM and the Extreme Science and Engineering Discovery Environment (XSEDE) allocations TG-AST20020 and TG-MCA94P018. XSEDE is supported by the National Science Foundation grant ACI-1053575. D.C. is a Ramon-Cajal Researcher and is supported by Ministerio de Ciencia, Innovación y Universidades (MICIU/FEDER) under research grant PID2021-122603NB-C21. N.M. acknowledges support from ISF grant 3061/21 and from BSF grant 2020302. C.H. is supported by NSF grant AAG-1911233, and NASA grants HST-AR-15800, HST-AR-16633, and HST-GO-16703. S.M., A.M., and P.M.  carried out this research under the auspices of the Science Internship Program at the University of California, Santa Cruz under the mentorship of C.S. 

We thank J.X. Prochaska, Joe Burchett, and Kirill Tchernyshyov for help finding and accessing observational CGM data, as well as David Koo, Sandra Faber, Farhanul Hasan, Doug Hellinger, Susan Kassin, Mia Bovill, and Benjamin Tufield for helpful discussions of this work in progress. The publicly available \code{enzo}, \code{yt}, and \code{trident} codes used in this work are the products of collaborative efforts by many independent scientists from numerous institutions around the world. Their commitment to open science has helped make this work possible.

\bibliographystyle{aasjournal}
\bibliography{refs}

\begin{appendix}

\begin{SidewaysFigure}
\includegraphics[clip,width=\linewidth]{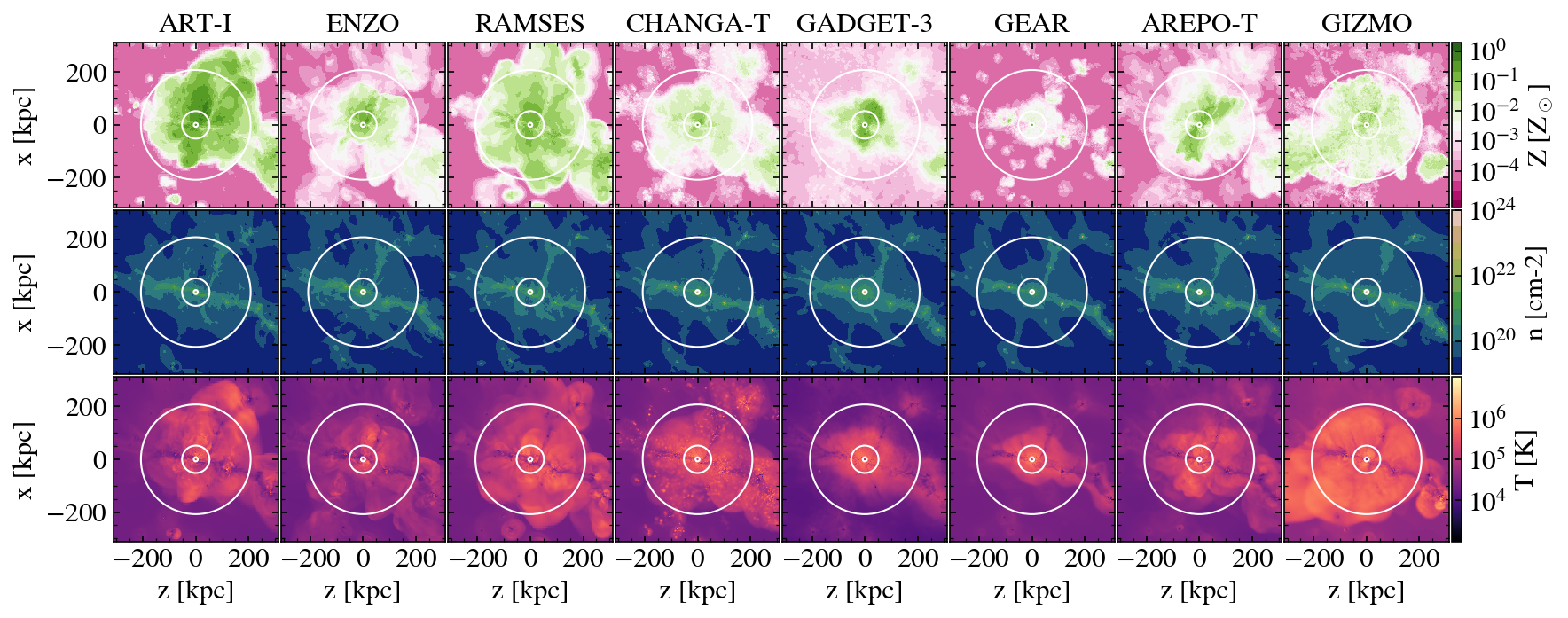}
\caption{Projection Plot at redshift \redshift{3}, identical to and aligned with Figure \ref{fig:projection_8codes_z3}, but at larger scale and not including radial velocity field. Here we show eight codes in three fields out to 6 times \Rvirtext. As before, inner and outer white circles represent 0.15 and 1.0 \Rvirtext, respectively (at this scale, the former appears point-like). Additional black circle represents the approximate simulation zoom-in region of 4.0 \Rvirtext. Rows (from top) are metallicity, number density, and temperature.}
\label{fig:wideprojection_8codes_z3}
\end{SidewaysFigure}

\section{Larger-scale cosmological context}\label{sec:appendix}
In Figure \ref{fig:wideprojection_8codes_z3} we show a copy of Figure \ref{fig:projection_8codes_z3} at much larger physical scale, now out to 6.0 \Rvirtext in each direction, with the approximate zoom-in region of 4.0 \Rvirtext outlined in white. There are two main effects visible in this figure. First, we show the full extent of the metal pollution of the IGM from each of the AGORA galaxies. While the biconical outflows are very visible on the small scale, at this scale the azimuthal differences become negligible. Instead, each code fills in a rough sphere of metals, with varying distances according to feedback strength. As a result of their fast, metal-rich outflows, \code{art-i} and \code{ramses} fill the whole volume out to 4.0 \Rvirtext at high metallicity close to solar values. \code{changa-t} and \code{gizmo} fill a similarly sized sphere, but at lower metallicities near $0.01 \zsun$. \code{enzo}, \code{gadget-3} and \code{arepo-t} fill out a smaller sphere, or only parts of it, leaving the biconical outflows somewhat more visible. Finally, \code{gear} remains fairly low metallicity out to large radii as commented on in Section \ref{sec:comparison}.

The second effect visible in this figure is the interconnection between the cool streams mentioned throughout the text and the larger-scale cosmic web. Intergalactic filaments are generally the source of these streams \citep[e.g.][]{birnboim_virial_2003,dekel_galaxy_2006}, and we see in the AGORA galaxies here that there are three major filaments entering the density and temperature pictures with roughly the same orientations as the ``streams'' mentioned in discussion of Figure \ref{fig:projection_8codes_z3}. While we commented in Section \ref{sec:comparison} that these sometimes mix before entering the galaxy or even before entering the halo, depending on fairly sensitive numerical effects, on this scale the same structures are always visible in all codes, due to the shared initial conditions. All codes contain high-temperature regions around their central galaxy, which have some overlap with their metal-rich spheres, however the exact temperature and size can vary. For example, \code{changa-t} in the temperature projection looks similar to \code{art-i} and \code{ramses}, and \code{gear} is more or less indistinguishable from \code{gadget-3} and \code{arepo-t}. Notably, we can see that on the IGM scale, the \code{gizmo} code is significantly hotter all the way out to 4.0 \Rvirtext than the others, even though in the CGM and galaxy it has similar dynamics to the other codes.

\end{appendix}


\end{document}